%% file: Haldane_arx.tex
\documentclass[aps,pra,reprint,showpacs,floatfix,superscriptaddress]{revtex4-1}

\usepackage{amssymb,amsmath,amstext}                
\usepackage{graphicx}                                               
\usepackage{epstopdf}                                               
\usepackage{color}                                                     
\usepackage{bm}                                                        
\usepackage{appendix}                                              
\usepackage[utf8]{inputenc}
\usepackage{bbold}
\usepackage{bbm}
\usepackage{latexsym}
\usepackage[T1]{fontenc}
\usepackage{xcolor}
\usepackage{braket}
\definecolor{lblue} {RGB}{51,71,158}
\usepackage[colorlinks=true,citecolor=blue,linkcolor=blue,urlcolor=blue]{hyperref}

\usepackage[normalem]{ulem}

\definecolor{darkgreen}{rgb}{0.13, 0.55, 0.13}

\newcommand{\tc}[1]{{\color{blue}{{#1}}}}

\newcommand{\new}[1]{\textcolor{black}{{#1}}}

\usepackage[english]{babel}

\def\beq{\begin{equation}}
\def\eeq{\end{equation}}
\def\ba{\begin{align}}
\def\enda{\end{align}}
\def\bi{\begin{itemize}}
\def\ei{\end{itemize}}

\usepackage{tabularx}

\begin{document}

\title[]{Quantum phases of dipolar bosons in one-dimensional optical lattices}

\author{Rebecca Kraus}
\affiliation{Theoretical Physics, Saarland University, Campus E2.6, D--66123 Saarbr\"ucken, Germany}
\author{Titas Chanda}
\affiliation{Institute of Theoretical Physics, Jagiellonian University in Krakow, ul. Lojasiewicza 11,
	30-348 Krak\'ow, Poland}
\affiliation{The Abdus Salam International Centre for Theoretical Physics (ICTP), Strada Costiera 11, 34151 Trieste, Italy}
\author{Jakub Zakrzewski}
\affiliation{Institute of Theoretical Physics, Jagiellonian University in Krakow, ul. Lojasiewicza 11,
	30-348 Krak\'ow, Poland}
\affiliation{Mark Kac Complex Systems Research Center,  Jagiellonian University in Krakow, \L{}ojasiewicza 11, 30-348 Krak\'ow, Poland}
\author{Giovanna Morigi}
\affiliation{Theoretical Physics, Saarland University, Campus E2.6, D--66123 Saarbr\"ucken, Germany}
\date{\today} 

\begin{abstract}
We theoretically analyze the phase diagram of a quantum gas of bosons that interact via repulsive dipolar interactions. The bosons are tightly confined by an optical lattice in a quasi one-dimensional geometry. In the single-band approximation, their dynamics is described by an extended Bose-Hubbard model where the relevant contributions of the dipolar interactions consist of density-density repulsion and correlated tunneling terms. We evaluate the phase diagram for unit density using numerical techniques based on the density-matrix renormalization group algorithm. Our results predict that correlated tunneling can significantly modify the parameter range of the topological insulator phase.  At vanishing values of the onsite interactions, moreover, correlated tunneling promotes the onset of a phase with a large number of low energy metastable configurations.
\end{abstract}

\maketitle
\section{Introduction}

Quantum gases of atoms and molecules in optical lattices are formidable platforms for studying the emergence of complex states of matter from the dynamics of the individual constituents, thanks to the experimental control of the characteristic length and energy scales \cite{bloch2008many,gross2017quantum}. One prominent example is the observation of the quantum phase transition between Mott-insulator and superfluid phases \cite{Fisher1989,Greiner2002}, demonstrating that these systems are versatile quantum simulators of the Bose-Hubbard model \cite{bloch2008many}. The most recent confinement of ultracold dipolar gases in optical lattices  \cite{Baier2016}  and the combination of optical lattices and cavity setups \cite{landig2016quantum} has permitted to study the interplay between short and long-range interactions in these settings. These experiments reported dynamics that can be encompassed by the so-called extended Bose-Hubbard models  \cite{Dutta2015}, where these interactions are described by additional terms of the Bose-Hubbard Hamiltonian \cite{sowinski2012dipolar,Habibian2013,caballero2015quantum}. 

In a lattice the effect of a two-body potential results in interaction terms, proportional to the onsite densities on both contributing lattice sites, as well as in so-called correlated tunneling terms, where hopping from site to site depends on the occupation of the neighboring sites \cite{sowinski2012dipolar,Dutta2015,cartarius:2017}. Detailed studies of the extended Bose-Hubbard model for dipolar gases typically included only the density-density interaction terms. These terms can induce density modulations and, in one dimension and at unit density, are responsible for the emergence of the so-called Haldane topological insulator, namely, an incompressible phase with a non-local order parameter \cite{batrouni2013competing,batrouni2014competing,Rossini2012,dalla2006hidden,kawaki2017phase}. 

Correlated tunneling is known from studies of superconductivity \cite{strack1993hubbard,hirsch1994inapplicability,amadon1996metallic} and quantum magnets \cite{schmidt2006single}. In quantum gases of bosons, at sufficient large dipolar interaction strengths it gives rise to pair condensation \cite{schmidt2006single,sowinski2012dipolar} and to superfluidity with complex order parameter \cite{jurgensen2015twisted}. Recent works showed that correlated tunneling is responsible for the emergence of superfluidity at large onsite repulsions, where one would instead expect insulating phases \cite{biedron2018extended,Kraus2020,suthar2020staggered}. 
\new{The effect of correlated tunneling for large densities in an one-dimensional lattice was studied in Ref.  \cite{Kraus2020,biedron2018extended} and its two-dimensional extension was examined in Ref. \cite{suthar2020staggered}.} \new{In particular, preliminary studies of the influence of the correlated tunneling on the existence of the Haldane insulator for a certain parameter choice was performed in \cite{biedron2018extended}.}

In this work we \new{perform an extensive characterization of} the effect of correlated tunneling on the ground state of dipolar gases in (quasi) one dimension \new{for unit density}, focusing in particular on the existence and properties of the Haldane insulator. For this purpose we numerically  determine the phase diagram of the extended Bose-Hubbard model in one dimension and at unit density. We focus on the parameter regime where the Haldane insulating phase was predicted in Refs. \cite{batrouni2013competing,batrouni2014competing,Rossini2012,dalla2006hidden,kawaki2017phase} and, differing from those works, we systematically include correlated tunneling into our model. Motivated by recent experiments with low dimensional dipolar gases in optical lattices \cite{Baier2016,dePaz:2013,Moses:2015,Covey:2016,Reichsoellner:2017,Je,bohn2017cold}, we take care of linking the coefficients of the extended Bose-Hubbard model with the experimental control parameters, in order to preserve the correct scaling between the coefficients across the phase diagram. The phase diagram is evaluated by means of the Density Matrix Renormalization Group (DMRG) approach \cite{white1992density, White_prb_1993, Orus_aop_2014, schollwock2011density} and of its version simulating the thermodynamic limit, here referred to as the infinite DMRG (iDMRG) \cite{schollwock2011density, McCulloch_2008,Crosswhite_2008, Kjall_PRB_2013}.

This paper is organized as follows. In Sec.~\ref{Sec:2} we introduce the model, the extended Bose-Hubbard model for bosons interacting via onsite repulsion, nearest-neighbor repulsive interactions and nearest-neighbor correlated tunnelings. We then discuss the connection between the coefficients of the extended Bose-Hubbard model and the experimental realizations in quasi one-dimensional geometries. In Sec.~\ref{Sec:3} we analyze the resulting ground-state phase diagram for unit density. The conclusions are drawn in Sec. \ref{Sec:4}. The appendices provide details on the numerical implementations.

\section{Extended Bose-Hubbard model}
\label{Sec:2}

The model at the basis of our analysis is the one-dimensional extended Bose-Hubbard Hamiltonian $\hat{H}_{\text{BH}}$, that reads \cite{sowinski2012dipolar,cartarius:2017}:
\begin{eqnarray}
\label{Eq:BH} 
\hat{H}_{\text{BH}}&=&-t\sum_{j=1}^{L-1}\left(\hat{a}^\dagger_j\hat{a}_{j+1}+{\rm H.c.}\right) +\frac{U}{2}\sum_{j=1}^L \hat{n}_j\left(\hat{n}_j-1\right)\\
&+&V\sum_{j=1}^{L-1}\hat{n}_{j}\hat{n}_{j+1}-T\sum_{j=1}^{L-1}\left[\hat{a}^\dagger_j\left( \hat{n}_j+\hat{n}_{j+1}\right)\hat{a}_{j+1}+{\rm H.c.}\right], \nonumber
\end{eqnarray}
where the first line is the standard Bose-Hubbard model and the second line is due to additional nearest-neighbor interactions. 
Here, $L$ is the number of sites, the operators  $\hat{a}_j$ and $\hat{a}^\dagger_j$ annihilate and create, respectively, a boson at site $j=1,\ldots,L$, with $\left[\hat{a}_{j}, \hat{a}_{l}^\dagger\right]=\delta_{j,l}$, and the operator $\hat n_j=\hat a_j^\dagger \hat a_j$ counts the bosons at site $j$. The coefficients are assumed to be real. Specifically, the tunneling rate $t$ describes the nearest-neighbor hopping, which promotes superfluidity, and $t>0$. The onsite repulsion $U$, $U>0$, penalizes multiple occupation of a single site. In the "standard" Bose-Hubbard model, as given by the first line of Eq.\ \eqref{Eq:BH},  the ratio $t/U$ controls the phase transition from superfluidity (SF) to Mott-Insulator (MI) at commensurate densities \cite{Fisher1989}.

The second line of Eq. \eqref{Eq:BH} contains the terms due to the dipolar interactions. The term proportional to  $V$ describes density-density interactions that favor the formation of density modulations in the repulsive, $V>0$, case \cite{Menotti2007}. The last term is responsible for tunneling processes that depend on the density of the neighboring sites and are scaled by the coefficient $T$.  Here, we have omitted a pair tunneling term and 4-site scattering terms, since the corresponding coefficients are of higher order in the Bose-Hubbard expansion \cite{dutta2011bose,biedron2018extended,Kraus2020}. Moreover, we have omitted terms beyond nearest neighbors. These additional terms can significantly modify the phase diagram for large values of $V$ \cite{Kraus2020}, but give rise to small corrections for the parameters considered in this work.

\subsection{Order parameters}
\label{Sec:2.1}

We characterize the ground-state phase diagram of  Hamiltonian \eqref{Eq:BH} by means of the observables that we detail in what follows. We first determine the ground state energy $E(N)$ for $N$ particle over $L$ lattice sites, with $N=L$. The so-called charge gap $\Delta_c$ corresponds to the energy required to create a particle-hole pair and is obtained after finding the ground-state energies for $N-1$ and $N+1$ bosons \cite{batrouni2013competing,batrouni2014competing}: 
\begin{align}
\Delta_c= E(N+1)+E(N-1)-2E(N)\,. \label{charge_gap}
\end{align}
Its non-vanishing value in the thermodynamic limit signals an insulating phase. An insulator is also characterized by a finite value of the so-called neutral gap $\Delta_n$, corresponding to the difference between the energy $E_{\text{ex}}(N)$ of the first excited state and the energy $E(N)$ of the ground state \cite{batrouni2013competing,batrouni2014competing}:
\begin{align}
\Delta_n=E_{\text{ex}}(N)-E(N) \label{neutral_gap} \, .
\end{align}
The first excited state is numerically found by determining the lowest energy state in the subspace orthogonal to the ground state, see  Appendix \ref{App:DMRG}. In the SF phase the neutral gap vanishes in the thermodynamic limit.

We note that in one dimension the SF phase is strictly-speaking a Luttinger liquid with exponent $K>2$ \cite{biedron2018extended,deng2013polar,batrouni2014competing,cazalilla2011one}, thus the off-diagonal correlations decay with the distance according to a power-law:
\begin{align}
	C_{SF}(r)=\left\langle \hat{a}_{j}^\dagger \hat{a}_{j+r} \right\rangle \sim r^{-1/2K}. \label{off_diag}
	\end{align}
In order to reveal modulations in the off-diagonal correlations, we calculate \new{the Fourier transform of the single-particle density matrix $M(q)$:}
	\begin{align}
	M(q)= \frac{1}{L^2}\sum_{i,j=1}^{L-1} e^{iq(i-j)} \left\langle \hat{a}_i^\dagger \hat{a}_j\right\rangle. \label{Mq}
	\end{align} 
 \new{Typically, in a standard SF the maximum component of $M(q)$ is at $q=0$.} The correlated tunneling, on the other hand, gives rise to effects that in one dimension are analogous to an effective change of the sign of the tunneling coefficient. Correspondingly, the Fourier transform \new{of the single-particle density matrix} can have a non-zero component at $q=\pi$. We dub  the corresponding ground state as staggered superfluid (SSF) phase \cite{Kraus2020}.

	\begin{table*}[t]
		\begin{tabular}{|p{4cm}|p{1.5cm}||p{1.5cm}|p{1.5cm}|p{1.5cm}|p{1.7cm}|p{1.5cm}|p{1.5cm}|p{1.5cm}|} 
			\hline
			\small Phase &  Acronym & \small{Charge gap} &\small{Neutral gap} &\small{Fourier trans.} &\small{Density modulation}& \small{String order} &\small{String order} & \small{Parity order}\\ 
			&   &\small{$\Delta_c$, Eq.\ \eqref{charge_gap} } &  \small{$\Delta_n$, Eq.\ \eqref{neutral_gap}}&   \small{$M(\pi)$, Eq.\ \eqref{Mq}}& \small{$S(\pi)$, Eq.\ \eqref{DW_orderparameter}}& \small{$\mathcal{O}_S(\rho)$, Eq.\ \eqref{eq:string}} &  \small{$\mathcal{O}_S(\left\langle \hat{n}_j\right\rangle )$, Eq.\ \eqref{eq:string}}&\small{$\mathcal{O}_P$, Eq.\ \eqref{eq:parity}}\\[3ex]
			\hline\hline
			\small{Mott Insulator} & \small{MI}      & $\neq 0$  & $\neq 0$&$= 0$&$=0$ & $=0$  &  $=0$&$\ne0$ \\
			\small{Charge Density Wave } &\small{CDW} & $\neq 0$ & $\neq 0$ &$= 0$& $\neq 0$ & $\neq 0$ & $=0$ &$\ne 0$\\
			\small{Haldane Insulator} &\small{HI}      & $\neq 0$& $\neq 0$ &$= 0$&$=0$ & $\neq 0$ & $\ne 0$ &$=0$\\
			\small{Lattice Superfluid} & \small{SF}  & $=0$ & $=0$ &$= 0$& $=0$ & $=0$ & $=0$ &$=0$\\
			\small{Lattice Supersolid} & \small{SS}     & $=0$ & $=0$& $= 0$& \small{$\neq 0 $} & $\neq0$& $=0$&$\neq0$  \\
			\small{Lattice staggered Superfluid} & \small{SSF}  & $=0$ & $=0$ &$\neq 0$&  \small{$=0$} & $=0$ &$=0$ &$=0$\\
			\small{Lattice staggered Supersolid} & \small{SSS}     & $=0$ & $=0$& $\neq 0$& $\neq 0 $ & $\neq0$ & $=0$ & $\neq0$  \\
			\hline
		\end{tabular}
		\caption{Table of the phases, of their acronyms, and of the corresponding values of the observables.}\label{Table:1}
	\end{table*}

Density modulated phases are revealed by properties of the local density-density correlations \cite{Rossini2012,berg2008rise}, whose Fourier transform is the structure form factor:
\begin{align}
S(k)=\frac{1}{L^2}\sum_{i,j}^{L-1}e^{ik(i-j)}\left\langle \hat{n}_i\hat{n}_j \right\rangle \label{DW_orderparameter} \,. 
\end{align}
For a two-site translational symmetry, $S(k)$ shows a finite peak at $k=\pi$. The phase is a Charge Density Wave (CDW) or lattice Supersolid (SS) depending on whether the density-modulated phase is incompressible or superfluid, respectively. The SS phase is a staggered supersolid (SSS) when $M(q)$ is finite and maximum at $q=\pi$.

The Haldane insulating phase (HI) is {gapped and characterized} by non-local spatial correlations in the density fluctuations \new{ $\delta \hat n_j$.} 
This is captured by the string order parameter $\mathcal{O}_S$ \cite{batrouni2014competing,dalla2006hidden, berg2008rise, batrouni2013competing,Rossini2012}:
\begin{align}
&\mathcal{O}_S=\lim_{r\rightarrow \infty}O_S(r) \label{eq:string}\\
&\text{with} \ \ O_S(r)=\left|\langle \delta \hat{n}_i e^{ \left(i\pi \sum_{k=i}^{i+r}\delta \hat{n}_k \right)} \delta \hat{n}_{i+r}\rangle \right| \nonumber \, . 
\end{align}
\new{The definition of the density fluctuation $\delta \hat n_j$ is important. When we consider the density fluctuations about the mean value $\rho$, namely,  $\delta\hat{n}_j=\hat n_j-\rho$, then we label the string order parameter by $\mathcal{O}_S(\rho)$. When instead  the density fluctuations are taken about the local mean occupation $\left\langle \hat{n}_j \right\rangle$, namely, $\delta \hat n_j(\left\langle \hat{n}_j \right\rangle )=\hat n_j-\left\langle \hat{n}_j \right\rangle $, then the corresponding string order parameter is given by  $\mathcal{O}_S(\left\langle \hat{n}_j \right\rangle )$. Both definitions give finite values within the HI phase. Instead, in the CDW phase $\mathcal{O}_S(\left\langle \hat{n}_j\right\rangle )$ vanishes, while $\mathcal{O}_S(\rho)$ is finite. Thus $\mathcal{O}_S(\left\langle \hat{n}_j\right\rangle )$ signals the HI phase.}
The HI phase can also be distinguished from other insulating phases by means of the parity order parameter  
\begin{align}
&\mathcal{O}_P=\lim_{r\rightarrow \infty}O_P(r) \label{eq:parity}\\
&\text{with} \ \ O_P(r)=\left|\langle e^{ \left(i\pi \sum_{k=i}^{i+r}\delta \hat{n}_k \right)}\rangle \right| \nonumber \,, 
\end{align}
which is finite in the MI and CDW phases, while it vanishes in the HI phase \new{independent of the definition of $\delta \hat{n}_j$.} 

The phases and the corresponding values of  order parameters are summarized in {Table~\ref{Table:1}}. 

Finally, we determine the von-Neumann entropy of the ground state for a lattice bipartition into two subsystems $A$ and $B$. Denoting the ground state by  $\ket{\psi_0}$, the von-Neumann (entanglement) entropy is defined as
\cite{amico2008entanglement,eisert2010colloquium,metlitski2011entanglement,frerot2016entanglement}
 \begin{align}
 S_{\text{vN}}=-\text{Tr}\left\lbrace \hat{\rho}_{B} \ln\left(\hat{\rho}_B \right) \right\rbrace \label{vonN} \ ,
 \end{align}
 where $\hat{\rho}_B= \text{Tr}_A\left\lbrace \ket{\psi_0}\bra{\psi_0} \right\rbrace$. 

\subsection{Bose-Hubbard coefficients}

The extended Bose-Hubbard model of Eq.~\eqref{Eq:BH} is a good approximation of the Hamiltonian describing the dynamics of dipolar atoms tightly confined by the lowest band of an optical lattice in a quasi one-dimensional geometry. The trapping potentials can be described by a potential of the form:
\begin{align}
\label{Eq:trap}
V_{\text{trap}}= \frac{m\omega^2}{2} \left(y^2 + z^2 \right)+ V_0\sin^2(\pi x/a)\  , 
\end{align}  
where $m$ is the atomic mass, $\omega$ is the frequency of the harmonic trap that confines the atomic motion along the $x$ direction, and $V_0$ is the depth of the optical lattice with periodicity $a$. The details of the derivation  of Eq. \eqref{Eq:BH}, starting from the full Hamiltonian of interacting atoms in the potential of Eq. \eqref{Eq:trap}, have been extensively reported, for instance, in Refs. \cite{cartarius:2017,Kraus2020}. These derivations allow one to link the Bose-Hubbard coefficients with the experimental parameters. 

In this work we set $V_0=8 E_R$ and $\omega= \sqrt{2V_{\text{har}}\pi^2/a^2m}$ with $V_{\text{har}}= 50E_R$ and $E_R=h^2/2m(2a)^2$ the recoil energy for a laser with wavelength $\lambda=2a$. The choice of $\omega$ warrants that the transverse motion is frozen out for the parameters that we consider. Since we keep the depth $V_0$ constant, the tunneling amplitude $t$ is fixed and finite. 

We sweep across the insulator-superfluid transition by varying the onsite interaction coefficient $U$. The latter results from the interplay between the van-der-Waals, contact potential $U_g(\mathbf{r})$ and the onsite contribution of the dipolar interaction $U_d(\mathbf{r})$:
\begin{eqnarray}
&&U_g(\mathbf{r})= g \delta^{(3)}\left(\mathbf{r}\right) \,,\\
&&U_d(\mathbf{r})= \frac{C_{dd}}{4\pi} \frac{1-3\cos^2(\theta)}{r^3}  \, . 
\end{eqnarray}
Here, $g  = 4\pi \hbar^2 a_s/m$ and is tuned by changing the scattering length, while the dipole-dipole potential is scaled by the coefficient $C_{dd}$ and $\theta$ denotes the angle between the dipoles and the interparticle distance vector $\mathbf{r}$. The other coefficients $V$ and $T$ are changed by varying the dipolar strength. 

\begin{figure}[h!]
 	\centering
 		\includegraphics[width=0.8\linewidth]{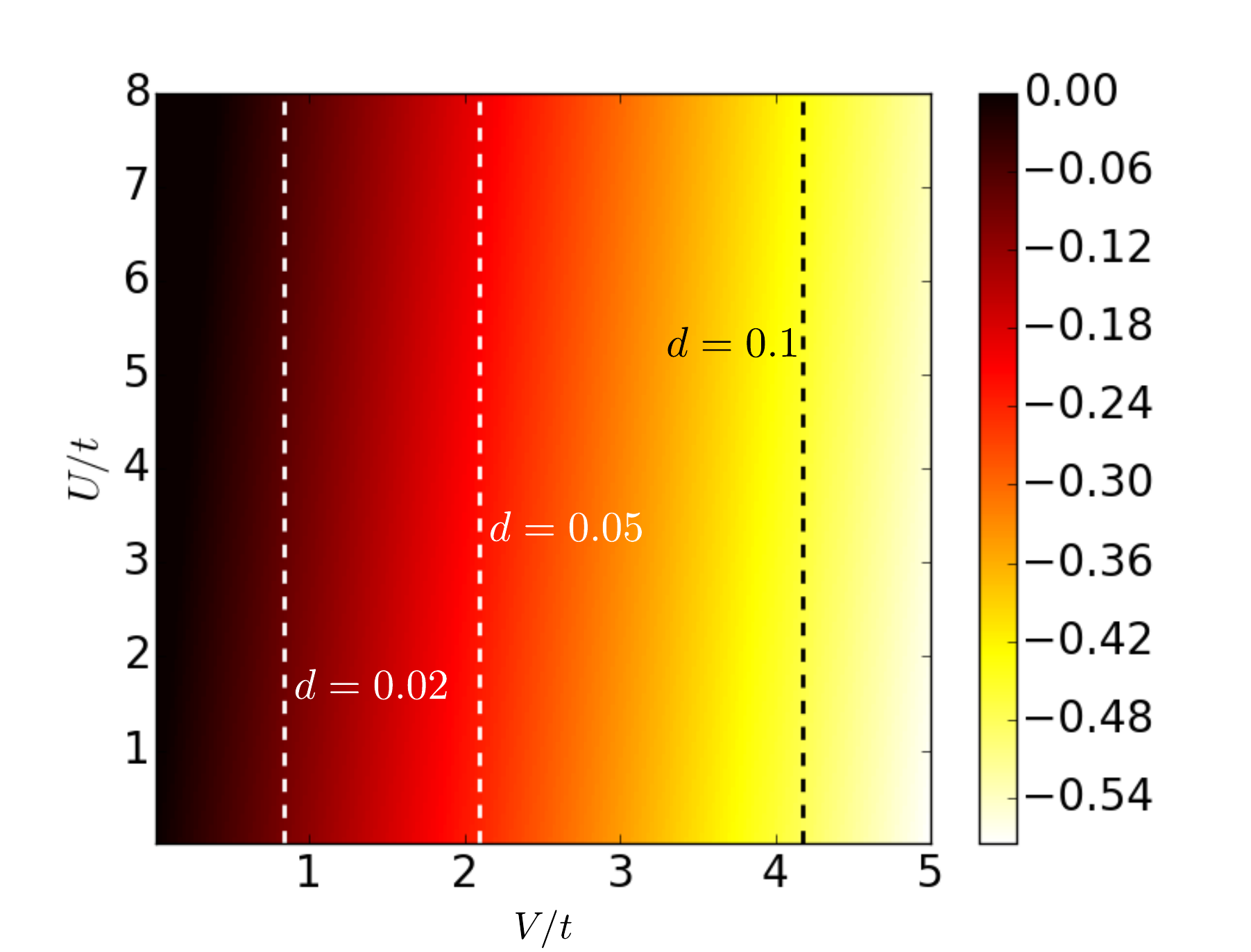}
 	\caption{Color plot of the correlated tunneling coefficient $|T|/t$ in the $V/t-U/t$-plane. All coefficients are in units of the tunneling rate, $t$. The black (white) dashed lines show the values of $V$ and $U$ at specific values of the dipolar interaction strength $d$. Note that $T\le 0$ across the phase diagram.} 
 	\label{fig1}
 \end{figure}

Figure~\ref{fig1} displays the absolute value of the correlated tunneling coefficient, $|T|$, as a function of the nearest-neighbor interaction, $V$, and of the onsite interaction, $U$. Both coefficients $|T|$ as well as $V$ increase with the dipole-dipole interaction strength, which is here reported in terms of the dimensionless parameter $d$ \cite{biedron2018extended, astrakharchik2007quantum}:
\begin{align}
d = \frac{mC_{dd}}{2\pi^3\hbar^2 a }\,.
\end{align}
{Note that} $T$ is negative for the parameters we consider and {it scales as} $|T|\sim V/10$.

\section{Ground-state phase diagram}
\label{Sec:3}

In this section we analyze the properties of the ground state of the extended Bose-Hubbard Hamiltonian in the $(U/t, V/t)$-plane and for \new{the} unit density. \new{We numerically determine the ground state on a finite lattice by means of DMRG  and extrapolate \tc{to} the thermodynamic limit of a given observable according to the procedure \cite{batrouni2014competing,Rossini2012}:
\begin{align}
\label{O:L}
\mathcal{O}\left(L\right)=\mathcal{O}(L\rightarrow \infty)+\text{A}/L + \text{B}/L^2  \ ,
\end{align}    
where $A$ and $B$ are constants, and $\mathcal{O}(L)$ stands for the observable at the lattice length $L$  \new{(see Appendix \ref{App:DMRG})}. } In our numerical simulations we take \new{$L= 64, 100, 128, 160$}. We identify the phase boundaries following the prescription given in Table~\ref{Table:1} for different observables. In this procedure we neglect the outer $L/4$ sites at both edges of the lattice in order to get rid of boundary effects and we evaluate the order parameters in the central part of the lattice, which consists of $r=L/2$ sites \cite{batrouni2014competing,Rossini2012} \new{(see Appendix \ref{App:DMRG})}. We compare these results with the phase diagram determined using iDMRG i.e., in the direct thermodynamic limit. Details of the implementations are provided in Appendix \ref{App:DMRG}.

\subsection{Phase diagram}

\begin{figure}[t]
	\centering
		\includegraphics[width=0.9\linewidth]{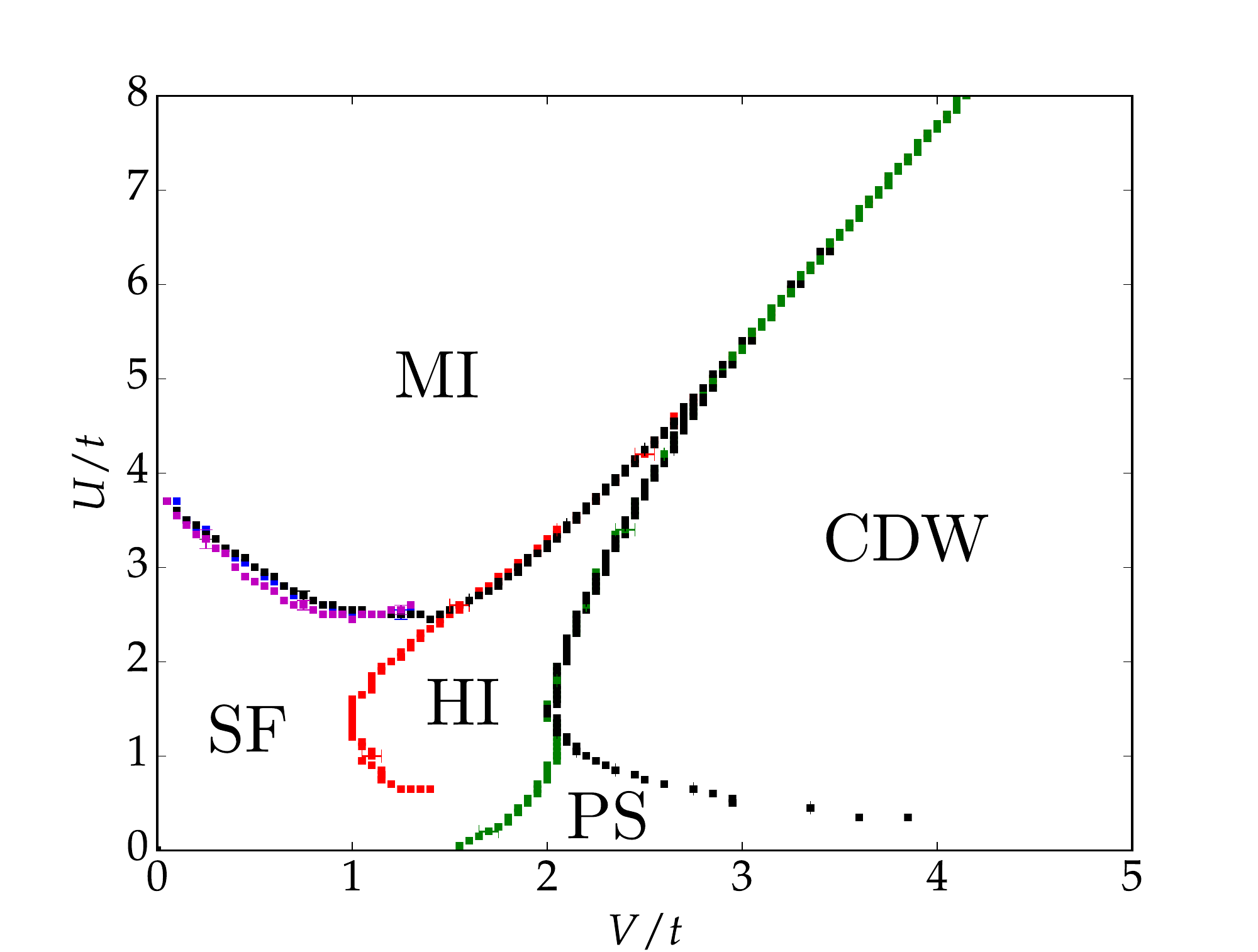}
	\caption{Phase diagrams in the $(U/t,V/t)$ plane for density $\rho=1$ obtained with DMRG on a finite lattice. The phases and boundaries are identified according to the behavior of the observables as in Table \ref{Table:1}.  The different colors indicate the parameters at which the corresponding observables vanish, namely, the neutral gap (magenta), the charge gap (blue), the parity (black), the string (red), and the density-wave (green) order parameters. The values are extrapolated to the thermodynamic limit from the data calculated with lattices of \new{$L=64,100,128,160$} sites (see text for details). We show few representative error bars. The error bars for each point are displayed in \new{Fig.~\ref{full_error}} in Appendix \ref{App:DMRG}.}
	\label{fig:2}
\end{figure}

The phase diagram for \new{the} density $\rho=1$ is shown in Fig.~\ref{fig:2} for a finite chain. The different colors indicate the phase boundaries predicted by (i) the charge gap (blue), (ii) the neutral gap (magenta), (iii) the string order parameter (red), (iv) the parity order parameter (black) and (v) the CDW order parameter (green). The boundaries are extracted following the procedure \new{described above, using} Eq.~\eqref{O:L}.

\begin{figure}[h]
	\centering
	\includegraphics[width=0.9\linewidth]{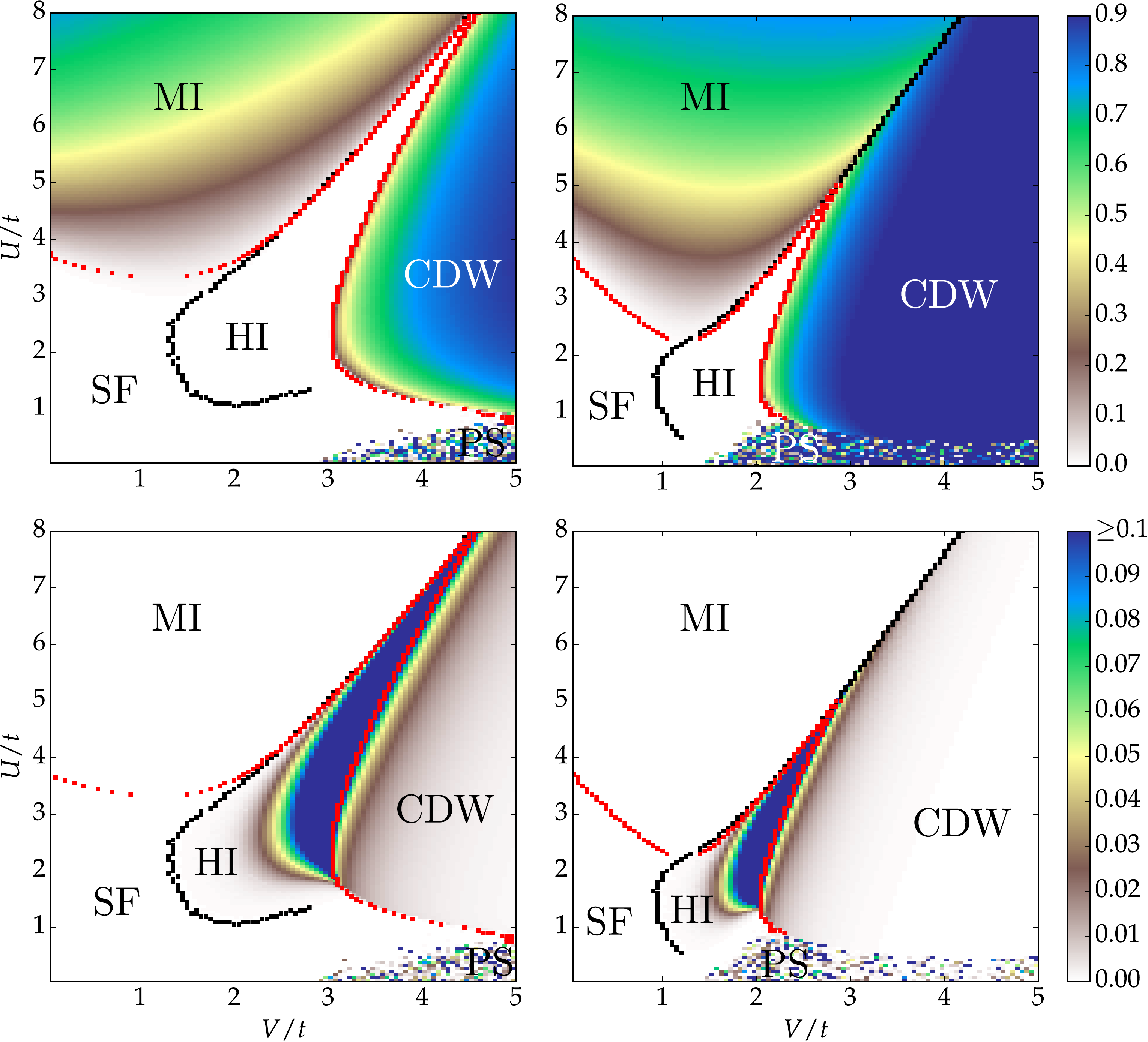} \\
	\caption{\new{String order parameter, $\mathcal{O}_S(\left\langle \hat{n}_i \right\rangle )$, \eqref{eq:string}, (lower panel) and parity order parameter $\mathcal{O}_P$, \eqref{eq:parity}, (upper panel)  in the $(U/t,V/t)$-plane obtained with iDMRG. The red (black) squares indicate the boundaries identified by vanishing parity, $\mathcal{O}_P$, (string, $\mathcal{O}_S(\rho)$) order parameter, respectively (see Appendix \ref{App:DMRG}). The left subplots show the order parameters for $T=0$, whereas the right subplots for $T\ne0$. }
	} 
	\label{String_inf}
\end{figure}

\new{In the considered parameter regime the phases  are SF, MI, HI, CDW, and a region which has the features of a phase separation (PS) and which will be discussed in Sec.\ \ref{PS}. We note that we do not find any staggered SF.}
\new{These findings are in agreement with the results obtained with infinite DMRG. Figure~\ref{String_inf} displays a color plot of the string order parameter $\mathcal{O}_S(\left\langle \hat{n}_i \right\rangle )$, (\ref{eq:string}) and of the parity order parameter $\mathcal{O}_P$, (\ref{eq:parity}), both obtained with iDMRG. For comparison, we also report the corresponding values obtained by setting $T=0$.}


Despite some similarities with the phase diagram found setting $T=0$  in Eq.\ \eqref{Eq:BH} \cite{batrouni2013competing, batrouni2014competing,Rossini2012}, nevertheless, there are also some striking differences. 
In the first place, for $T\neq 0$ the HI phase  occupies a smaller area in parameter space. {This confirms the observation in Ref.\ \cite{biedron2018extended}}. \new{In general, correlated tunneling stabilizes the MI and CDW phases in the parameter space, while the size of SF and HI phases are substantially reduced.} \new{Moreover, the HI phase seems to stretch down to smaller values of $U/t$ and $V/t$.}  \new{We note that we cannot determine the phase boundaries for small $U/t$ and around $1.5 \lesssim V/t \lesssim 2$} \new{because in this region the error bars are large.}


\subsection{The von-Neumann entropy}

\begin{figure}[h]
	\centering
	\includegraphics[width=0.8\linewidth]{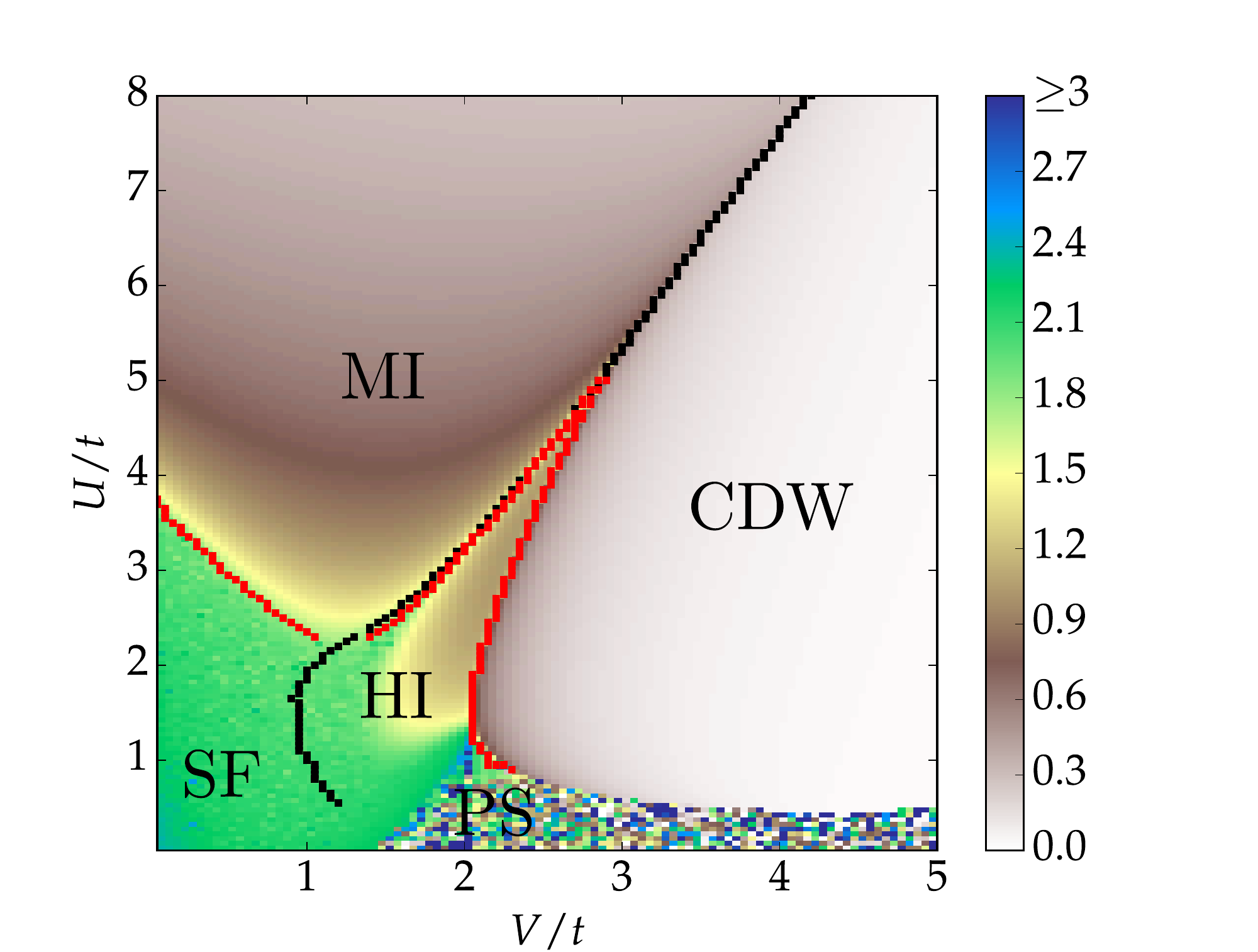} \\	
	\includegraphics[width=0.8\linewidth]{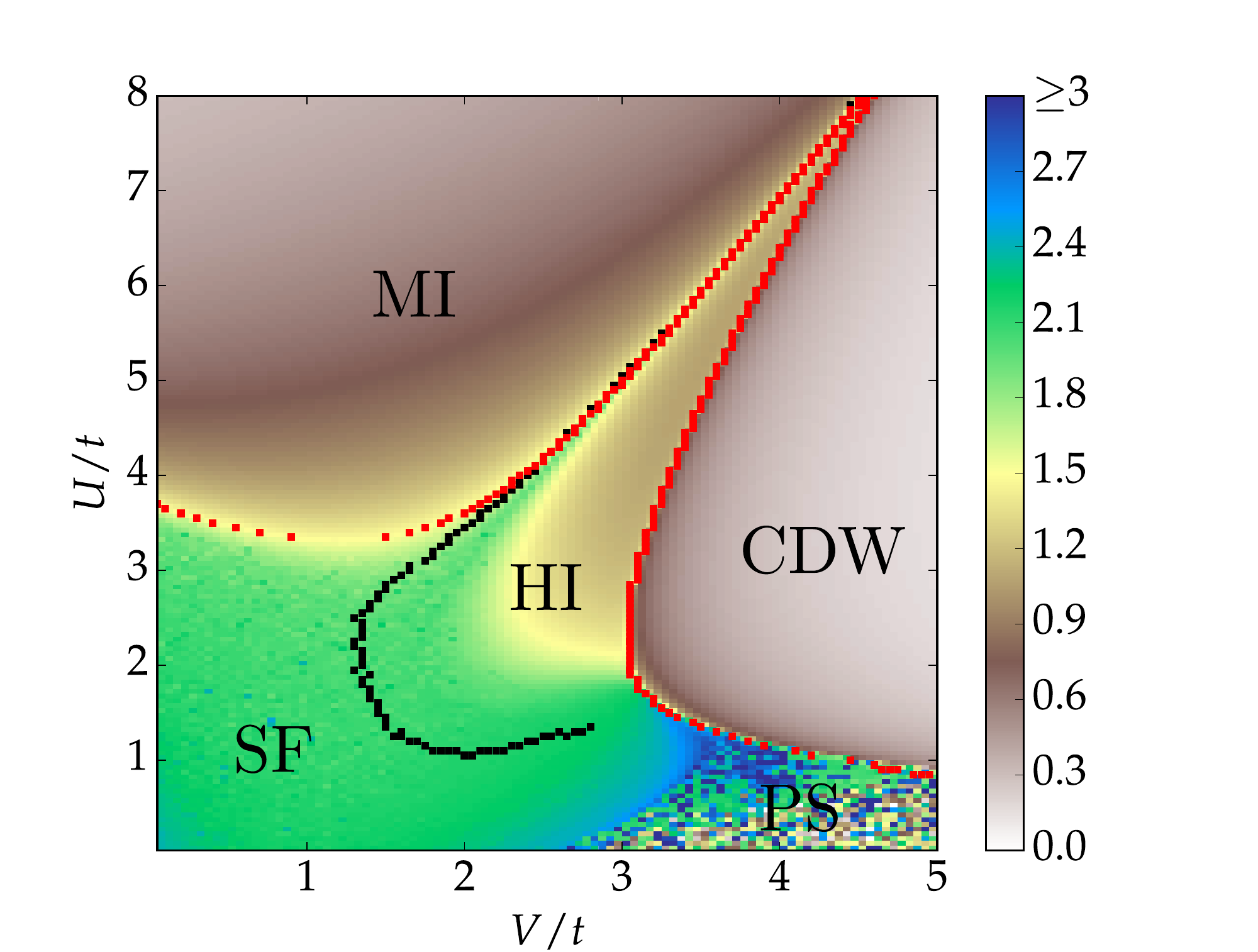}
	\caption{\new{Color plot of the von-Neumann entropy, Eq.~\ref{vonN}, in the $(U/t,V/t)$-plane using iDMRG. The squares indicate the boundaries identified using iDMRG and correspond to the values where the string $\mathcal{O}_{S}(\rho)$ (black) {and/or parity $\mathcal{O}_P$ (red)} order parameters vanish. The upper Subplot shows the von-Neumann entropy for $T\ne 0$, whereas the lower subplot depicts the von-Neumann entropy for $T=0$.}} 
	\label{vonNeu}
\end{figure}

\begin{figure}[h]
	\centering
		\includegraphics[width=0.9\linewidth]{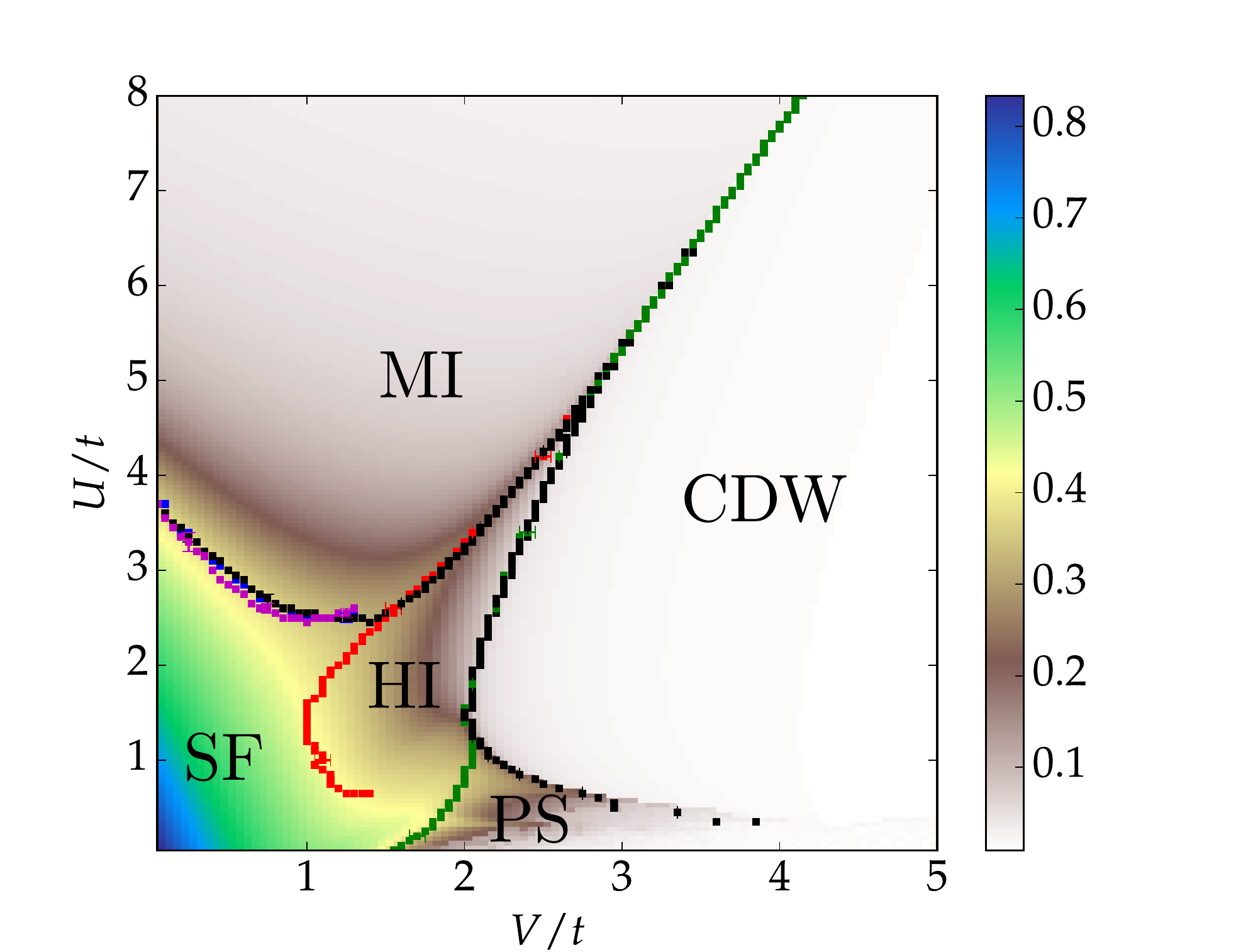}
	\caption{Fourier transform of \new{the single-particle density matrix} $M(q)$ at $q=0$,~\eqref{Mq}, in the $(U/t,V/t)$-plane. The data have been determined using DMRG  on a lattice with $L=100$. The different lines correspond to the phase boundaries identified by means of \new{the neutral gap (magenta), charge gap (blue),} parity (black), string (red), and density-wave (green) order parameters. We remark that everywhere $M(q)$ is maximum at $q=0$. In particular, we do not find staggered SF in the displayed parameter region.} 
	\label{Fig:6a}
\end{figure}

The color plots in Fig.~\ref{vonNeu} report  the von-Neumann entropy, $S_{\text{vN}}$ \eqref{vonN},  across the phase diagram and calculated by means of iDMRG. The von-Neumann entropy sheds light on the spatial decay of correlations. Comparison with the plot of the Fourier transform of the \new{single-particle density matrix}, Fig.\ \ref{Fig:6a}, shows that part of the region where $S_{\rm vN}$ is maximal overlaps with the SF domain. Like $M(q)$, the von-Neumann entropy decays slowly to zero when increasing $U/t$ at small values of $V/t$, sweeping across the SF-MI phase transition.

For small $U/t$ and for \new{$V/t\gtrsim 1.5$} $S_{\rm vN}$ undergoes strong fluctuations from point to point. We associate this behavior with the phase separation where the convergence of DMRG is doubtful. Comparing this region with the one at $T=0$, \new{lower panel of Fig.~\ref{vonNeu},} we observe that for $T\neq 0$ it appears at significantly lower values of $V/t$.

\begin{figure}[h]
	\centering
			\includegraphics[width=0.8\linewidth]{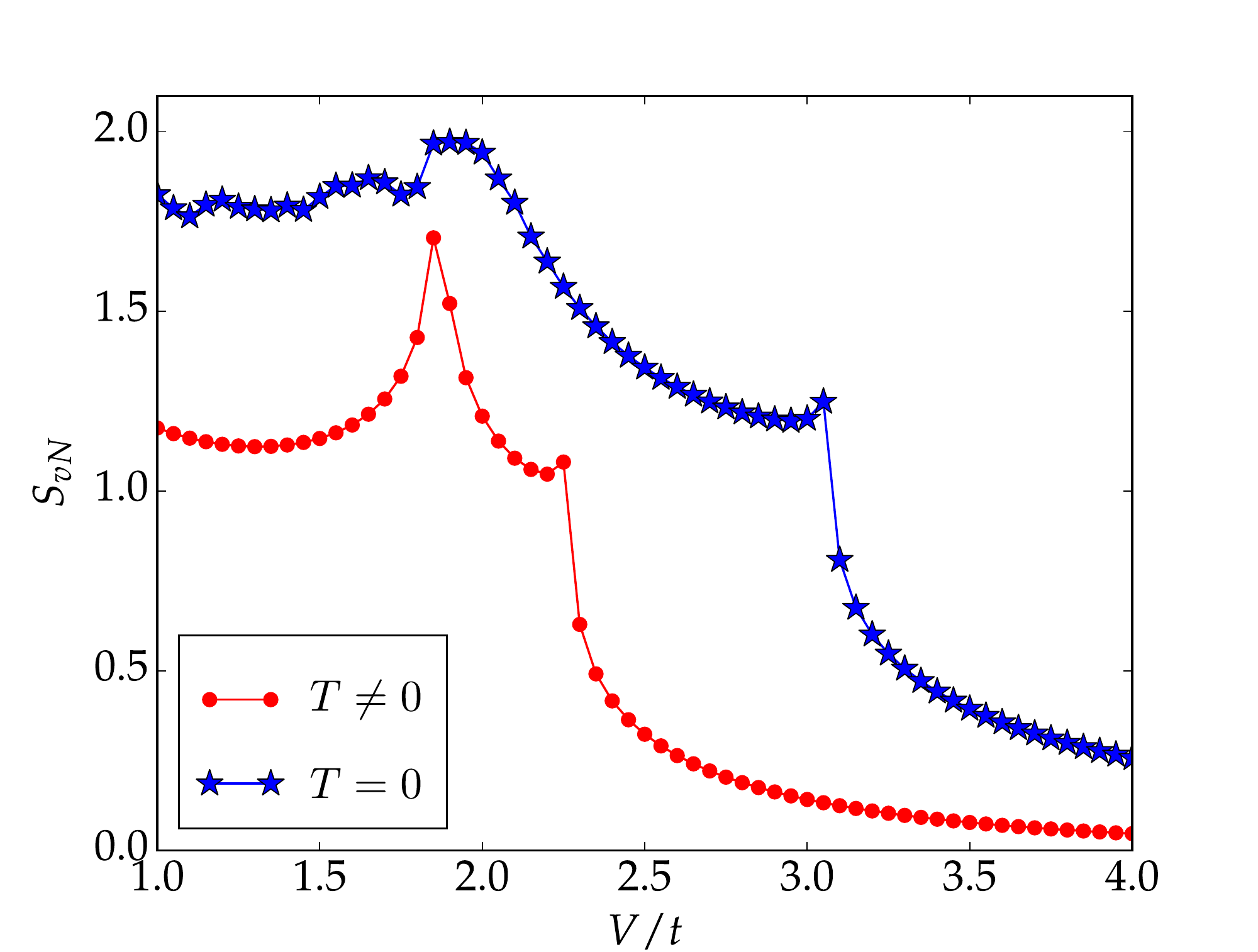}
	\caption{The von-Neumann entropy for fixed value of $U/t =3$ and as a function of the nearest-neighbor interaction strength $V$ in units of the tunneling $t$. The curve is a cut of the color plot in Fig.~\ref{vonNeu} calculated by means of  iDMRG.	
\new{We note that the noisy behaviour at the left part of the blue curve is within the SF phase, where the iDMRG for a large bond dimension is hard to converge.}} 
	\label{entanglement}
\end{figure}

Figure~\ref{entanglement} displays $S_{\rm vN}$ as a function of $V/t$ at fixed ratio $U/t$, in the part of the phase diagram where the phases are insulating. Starting from the MI phase we observe peaks when crossing the MI-HI and  the HI-CDW transitions, which we discuss in detail in the following.

\begin{figure}[h]
	\centering
		\includegraphics[width=0.8\linewidth]{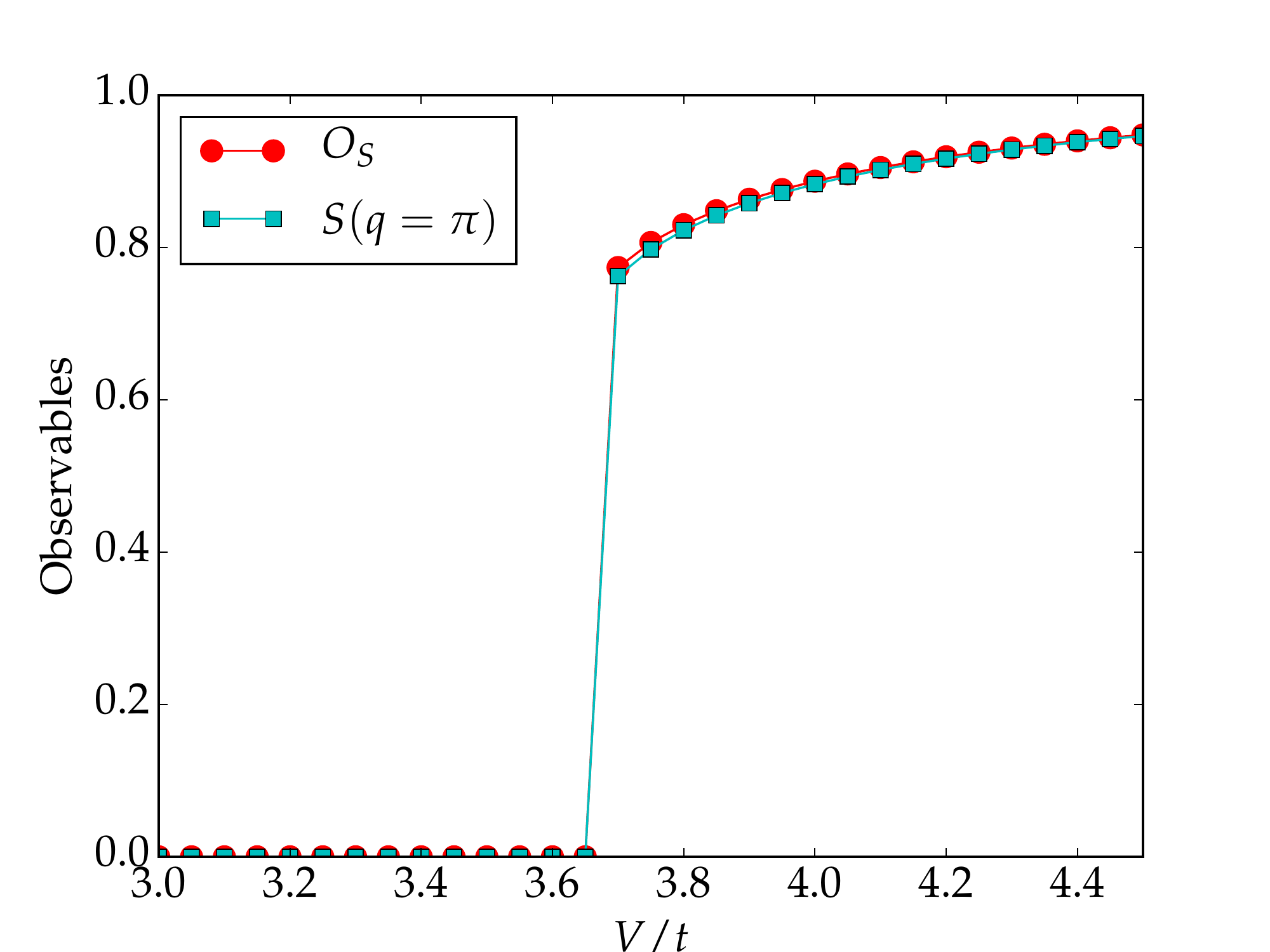}
	\caption{String and density-wave order parameter as a function of the nearest-neighbor interaction $V$ in units of $t$. The data have been calculated for finite $U/t=7$ by means of finite DMRG and extrapolated to the thermodynamic limit. Both order parameters are discontinuous at the MI-CDW transition, signaling a first order phase transition.} 
	\label{fig:3c}
\end{figure}

\subsection{{MI-HI-CDW transitions}}

In Fig.~\ref{fig:2} we observe a direct transition from the MI to the CDW phase at sufficiently high values of $U/t$. Fig.~\ref{fig:3c} shows that string \tc{($\mathcal{O}_S(\rho)$)} and density-wave order \tc{($S(q=\pi)$)} parameters are discontinuous at the transition point, indicating a first-order phase transition. \new{Here the string and density-wave order parameter agree almost exactly, since in the limit $U/t$ large the MI and CDW can be described by trivial Fock states, which lead to the same value of the string and density-wave order parameter in the thermodynamic limit.}

At smaller values of the ratio $U/t$ the HI phase separates the MI from the CDW phase.  The peaks in the profile of the von-Neumann entropy in Fig.~\ref{entanglement} suggest that the phase transitions at the MI-HI and   at the HI-CDW transitions are continuous (of second order). This is corroborated by the behavior of the neutral gap at the MI-HI and at the HI-CDW transitions. The HI phase corresponds to the interval where the energy gaps and the string order parameter possess finite values, while both parity and density-wave order parameter vanish. The finite value of the string order parameter and the vanishing parity order parameter demonstrate the topological nature of the HI phase. 

\begin{figure}[h]
	\centering
		\includegraphics[width=0.8\linewidth]{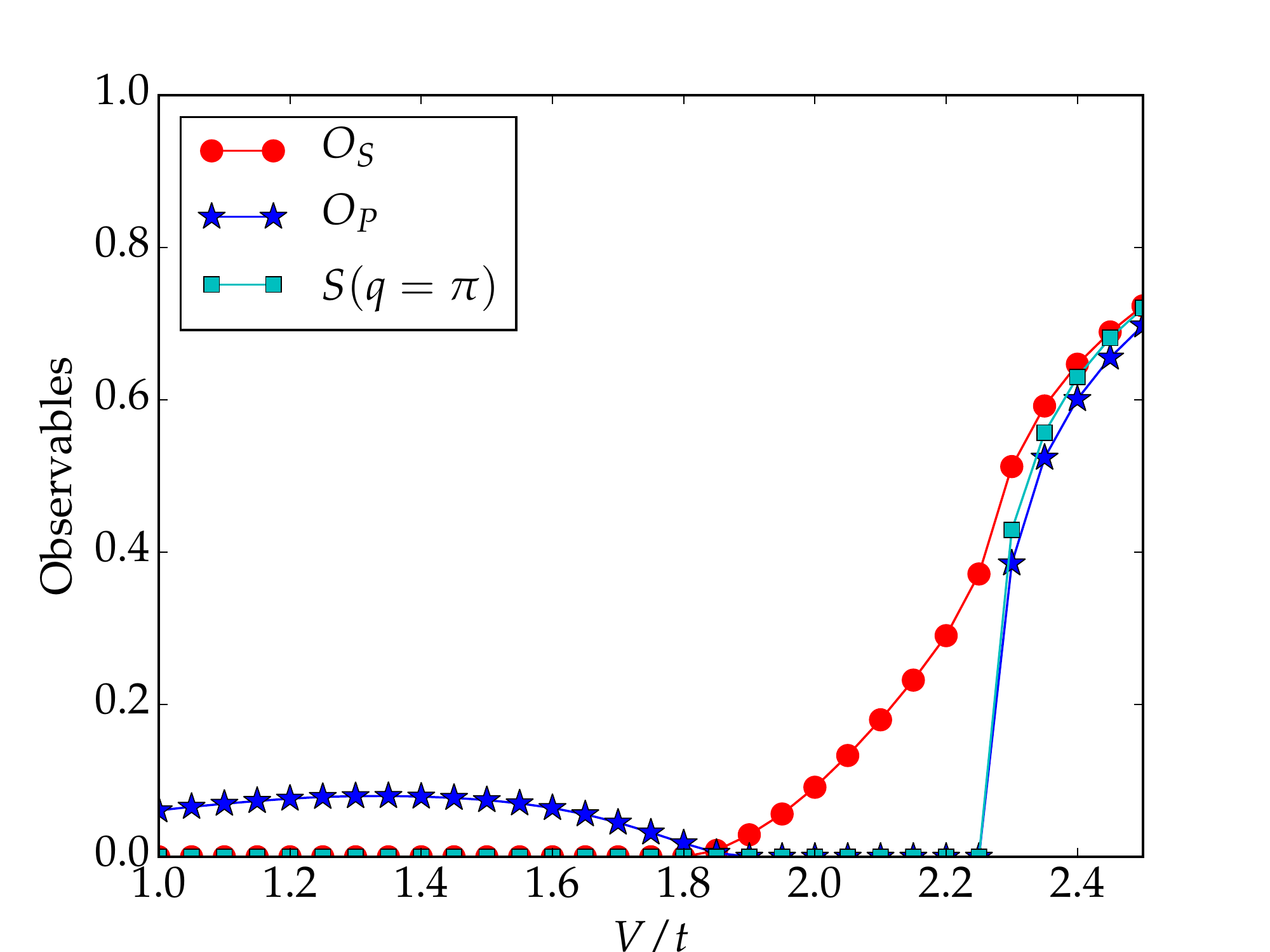} \\
		\includegraphics[width=0.8\linewidth]{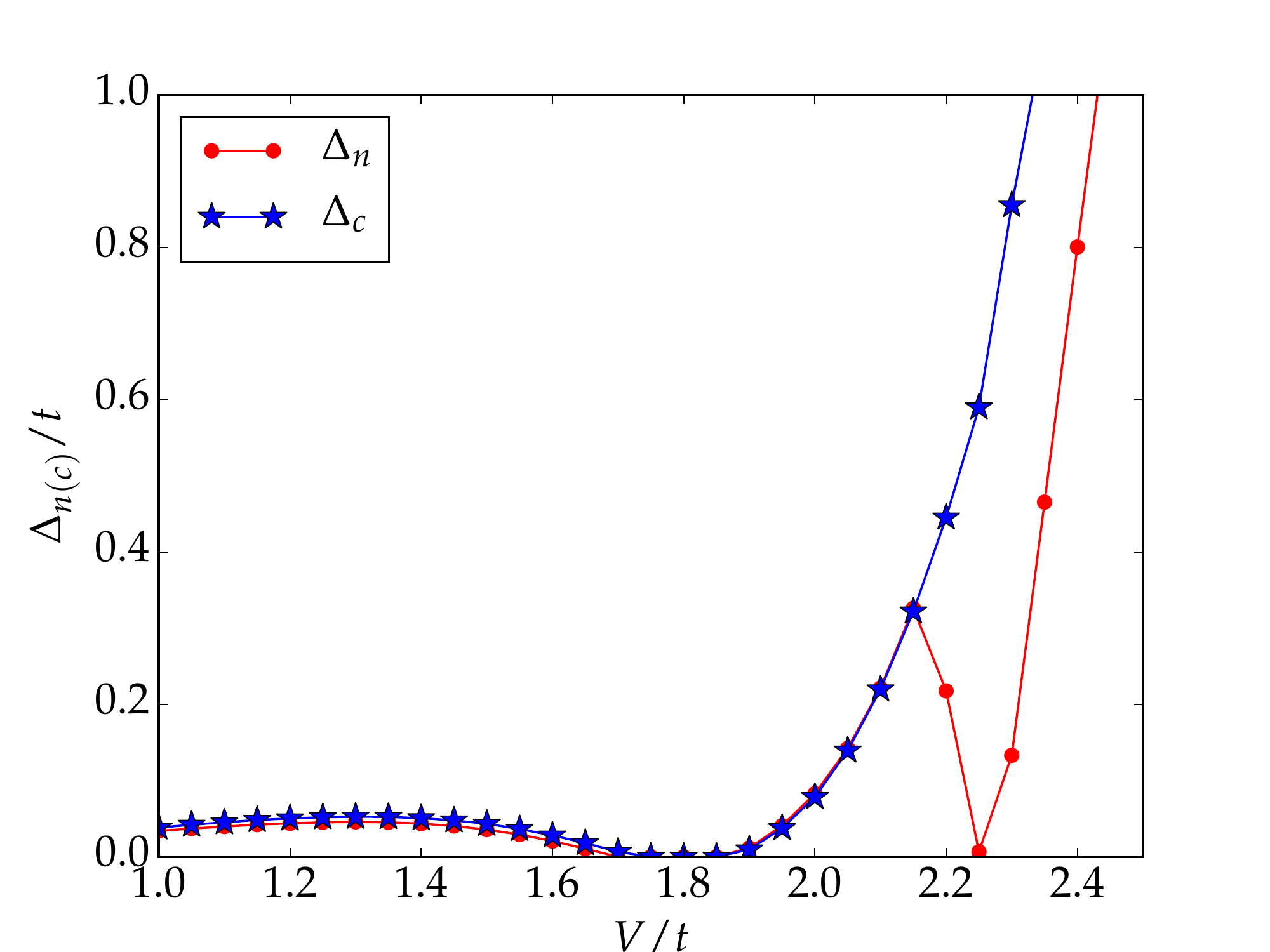}
	\caption{\new{Different o}bservables as a function of the nearest-neighbor interaction $V$ in the units of $t$ for finite $U/t=3$ calculated by means of finite DMRG and extrapolated to the thermodynamic limit. 
	Upper panel: string, parity, and density-wave order parameters. Lower panel: neutral and charge gaps. The color code is reported in the insets. We here take  a smaller value of the ratio $U/t$ with respect to the one of the corresponding Figure in Ref.\  \cite{batrouni2014competing}, since the phase boundaries for $T \ne  0 $ are shifted to smaller values of $U/t$ and $V/t$ with respect to the one for $T=0$ (see also Fig.~\ref{entanglement}).} 
	\label{fig:3}
\end{figure}

The neutral and charge gaps are displayed in the lower panel of Fig.~\ref{fig:3}  for $U/t=3$ as a function of $V/t$. For small $V/t$ the neutral and charge gaps are finite, corresponding to the MI phase. For a larger value of $V/t$  the gaps shrink to zero indicating the continuous transition to the HI phase. This agrees with the results for the $T=0$ case (no correlated tunneling) \cite{batrouni2014competing, berg2008rise, Ejima2015,ejima2014spectral}, where vanishing gaps both in the charge and neutral sectors signal a second order phase transition with central charge $c=1$ \cite{berg2008rise,ejima2014spectral,Ejima2015}. At the transition separating the HI and the CDW (symmetry-broken) phase the neutral gap vanishes, while the charge gap remains finite. This is as in the $T=0$ case, where the transition is of Ising type with central charge $c=1/2$ \cite{berg2008rise,amico2010hidden,ejima2014spectral, Ejima2015} and the quantum critical point is topological \cite{Fraxanet2021}. In the CDW phase the density-wave order parameter reaches a finite value, see the upper panel of Fig. \ref{fig:3}.

\subsection{Phase separation} \label{PS}

We finally discuss the parameter region at large $V/t$ but small $U/t$, where  the von-Neumann entropy has large fluctuations from point to point. We denote this regime by phase separation. Here, we find that the ground state of the canonical ensemble consists of a mixture of two or more phases. This feature can be revealed by inspecting the site occupation and its variance across the lattice. It can also be captured by the chemical potential as a function of the density $\rho$ \cite{batrouni2014competing,jakub2013}. In fact, in the grand-canonical ensemble the phase at unit density is unstable and the density is a discontinuous function of the chemical potential \cite{batrouni2014competing}. 

\begin{figure}[h]
	\centering
		\includegraphics[width=0.8\linewidth]{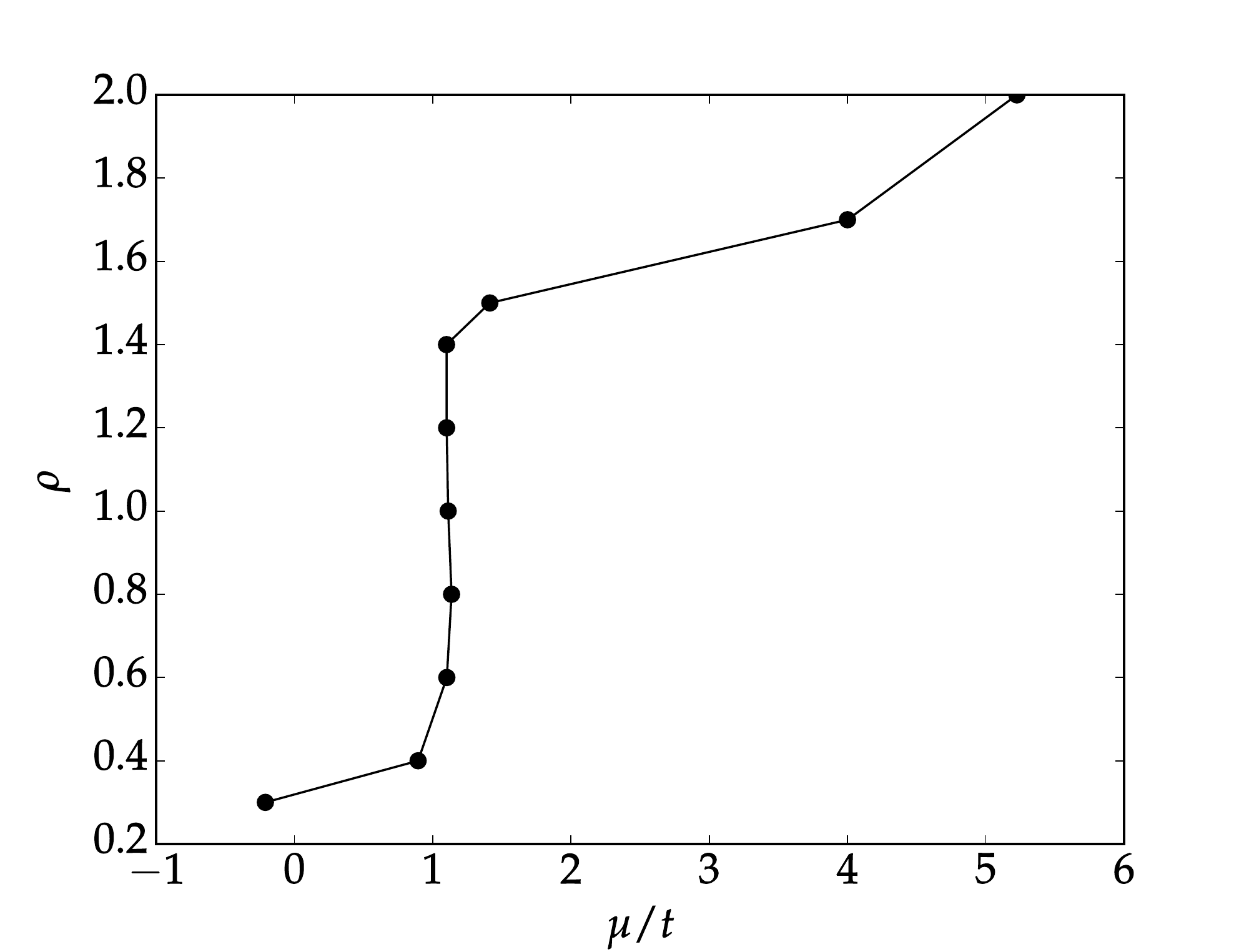}
	\caption{Density $\rho$ as function of the chemical potential $\mu$ (in units of $t$) for $U=0.5t$ and $V=4t$. The chemical potential is calculated according to Eq. (\ref{chem_pot}) by means of DMRG in a finite lattice with $L=20$.} 
	\label{Fig:5}
\end{figure}

In order to analyze the phase-separation region in the canonical ensemble, we calculate the density $\rho= N/L$ as a function of the chemical potential $\mu$, which we find by means of the formula \cite{batrouni2014competing}
\begin{align}
\mu(N) \approx E(N+1)-E(N) \ .  \label{chem_pot}
\end{align}
 Figure~\ref{Fig:5} displays $\rho$ as a function of $\mu$ for $(U/t,V/t)=(0.5,4)$ within the phase-separation region. The behavior suggests a hysteresis, which signals a discontinuous transition. 

\begin{figure}[h]
	\centering
		\includegraphics[width=0.9\linewidth]{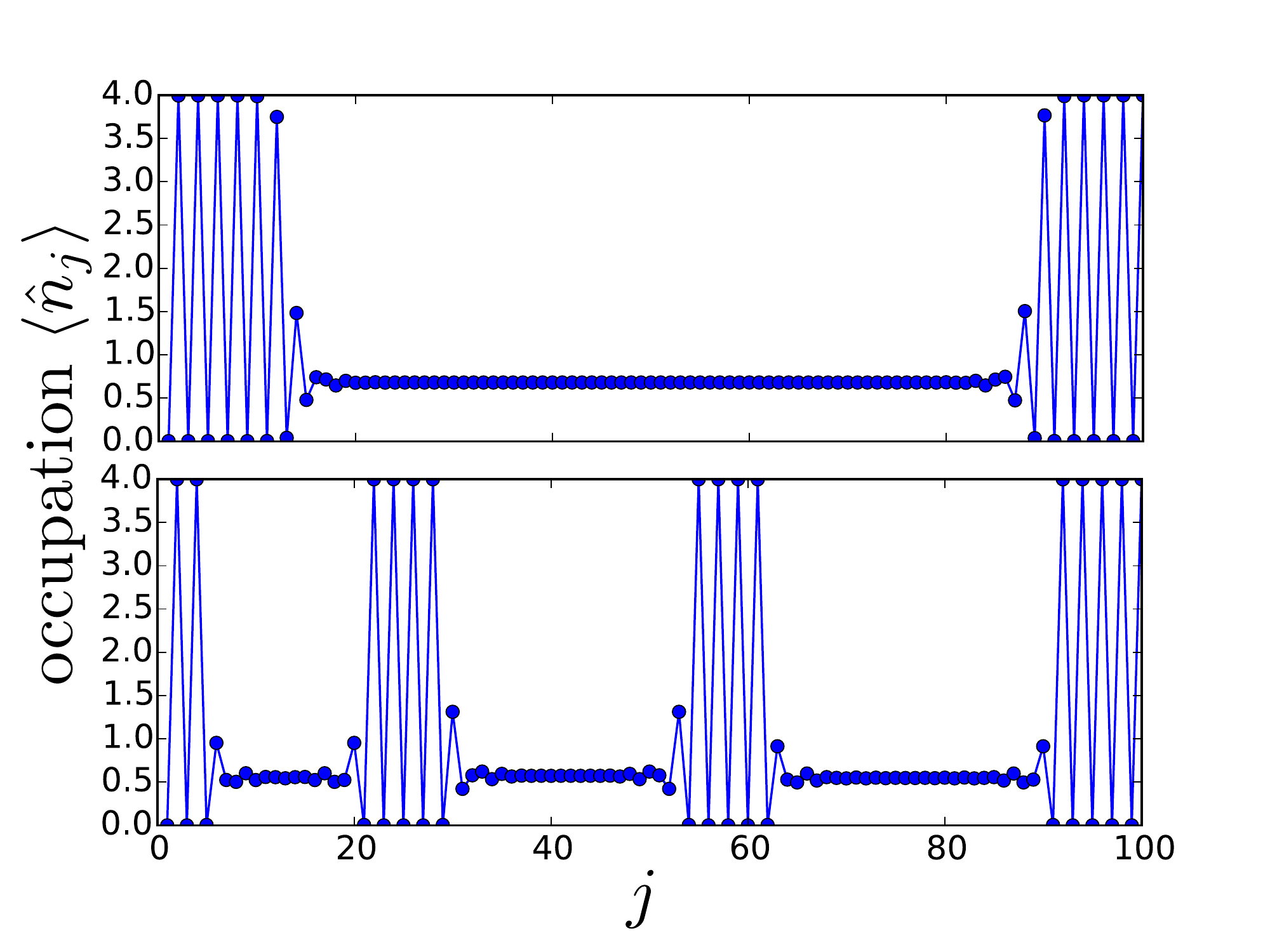}
	\caption{Typical metastable configurations in the phase separation regime. Occupation $\left\langle \hat{n}_j\right\rangle$ as function of the lattice site $j$ calculated by means of DMRG on a lattice with $L=100$ and for $U/t= 0.15$ and $V/t=2.2$ (upper panel), $U/t= 0.15$ and $V/t=2.8$ (lower panel). }
	\label{Fig:6}
\end{figure}

The phase separation region for $T=0$ has been recently extensively analyzed in  \cite{kottmann2021supersolid}. Correlated tunneling shifts the appearance of this phase to lower values of $V/t$ and possibly increases the number of metastable configurations. Figure~\ref{Fig:6} displays some of the metastable configurations we find, corresponding to CDW clusters separated by SF regions. Configurations like the one in the upper panel have been reported in \cite{batrouni2014competing}. The configuration in the lower panel, instead, seems to be stable due the presence of correlated tunneling.

\subsection{Discussion}

\begin{figure}[h]
	\centering
		\includegraphics[width=0.9\linewidth]{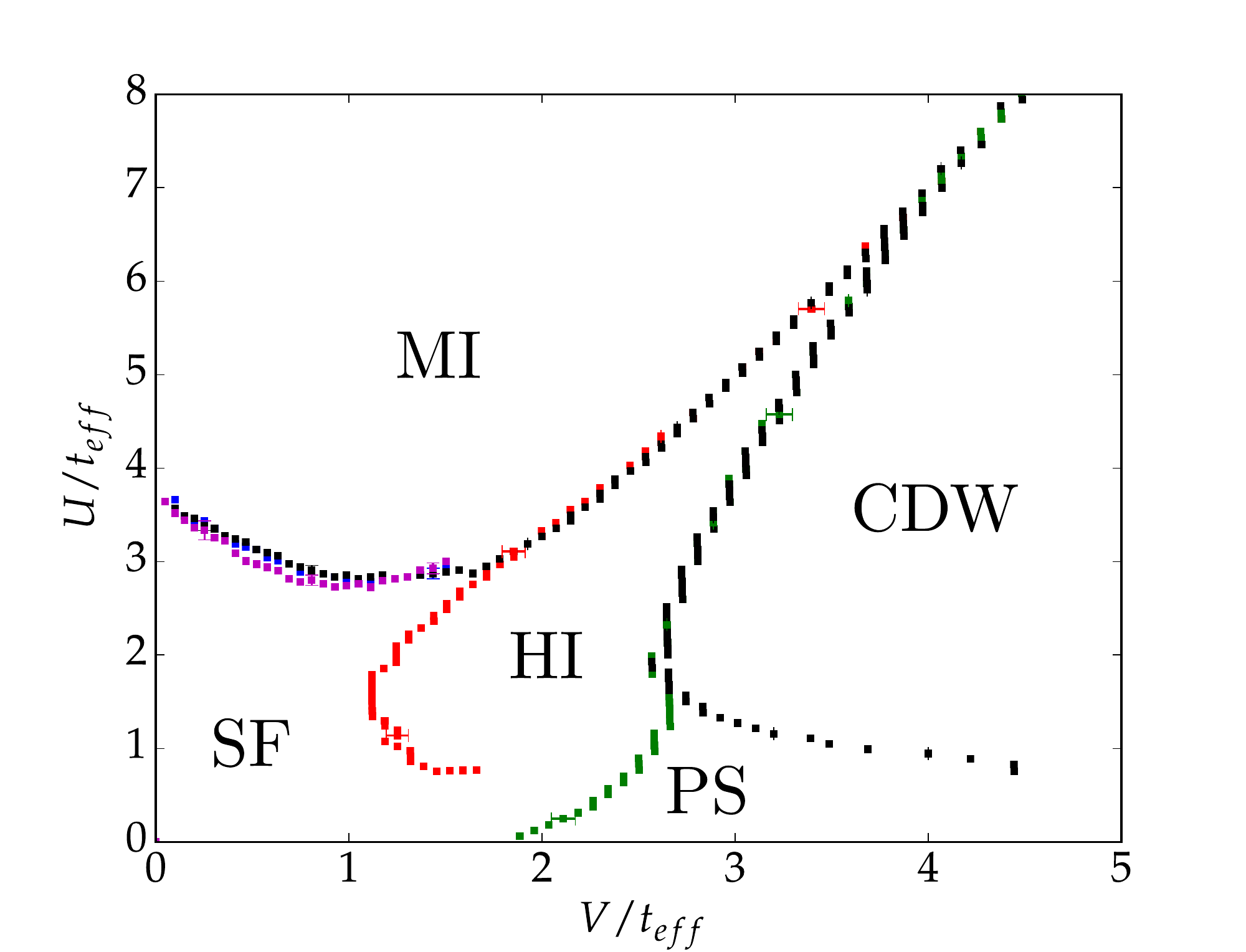}
	\caption{Color plot of the phase diagram in the $U-V$ plane. The data are the same as in Fig. \ref{fig:2}, the axes are here rescaled by the effective tunneling amplitude $t_{\rm eff}=t+T$, see text. } 
	\label{figapp}
\end{figure}

In previous works some of us showed that the effect of correlated tunneling on the ground-state phase diagram can be partially captured by an effective model. In this effective model correlated tunneling and single-particle hopping are replaced in Eq. \eqref{Eq:BH} by a single hopping term with effective tunneling coefficient $t_{\rm eff}=t+T(2\rho-1)$ \cite{Kraus2020,suthar2020staggered}. This coefficient can vanish, giving rise to an effective atomic limit which agrees with numerical results obtained with the full model \cite{Kraus2020,suthar2020staggered}. We have verified that, for the parameters we consider, $t_{\rm eff}$ is always finite. Fig.~\ref{figapp} displays the same data as in Fig. \ref{fig:2}, but with the axes now rescaled by $t_{\rm eff}$: The  rescaled phase boundaries SF-MI and MI-HI-CDW are in good agreement with the phase diagram at $T=0$ (c.f. Fig.~\ref{entanglement}(b)) \cite{batrouni2014competing}, suggesting that the effect of correlated tunneling on the size of the insulating phase could be captured by this  effective description.

\section{Conclusion}
\label{Sec:4}

We have analyzed the ground-state phase diagram of the extended Bose-Hubbard model in one dimension and unit density, describing a gas of dipolar bosons in an optical lattice and in a quasi one-dimensional geometry. 
\new{With respect to previous studies,} \new{in this work we have performed a systematic characterization of the effect of correlated tunneling on the phase diagram, focusing in particular on the parameter regime of the topological Haldane insulator.}

For the considered parameter space correlated tunneling plays a relevant role in determining the essential features of the phase diagram. By comparing with the phase diagrams calculated setting $T=0$  \cite{batrouni2013competing,batrouni2014competing,Rossini2012,berg2008rise}, we find that correlated tunneling tends to stabilize the insulating phases and to shrink the parameter region where the Haldane insulator is found. Moreover, correlated tunneling promotes the onset of the phase-separation regime also at relatively low values of the dipolar interactions, giving rise to a large number of low-energy metastable configurations.  \new{Future work will analyses relaxation after quenches. In fact, the Bose-Hubbard model with correlated tunneling exhibits several analogies with constrained models, which are known to give rise to a rich prethermalization dynamics \cite{nuske2020metastable,bluvstein2021controlling,Ney:2022}.} 

This study shows that correlated tunneling gives rise to correlations which are only partially captured by the observables typically employed for characterizing the phase diagram. These correlations might be also important at fractional filling. For instance, they might affect the properties of the Fibonacci anyonic excitations expected  at $\rho=3/2$ for low tunneling rates \cite{Duric17}.

\begin{acknowledgments}
The authors are grateful to Benoit Gremaud, and Luis Santos for discussions and especially to George Batrouni for helpful comments. RK and GM acknowledge support by the Deutsche Forschungsgemeinschaft (DFG, German Research Foundation) via the CRC-TRR 306 ``QuCoLiMa'', Project-ID No. 429529648, and by the priority program No. 1929 "GiRyd'''. We also thank funding by the German Ministry of Education and Research (BMBF) via the QuantERA project NAQUAS. Project NAQUAS has received funding from the QuantERA ERA-NET Cofund in Quantum Technologies implemented within the European Union's Horizon 2020 program. TC and JZ thank the support of  PL-Grid Infrastructure and the National Science Centre (Poland) under project Opus 2019/35/B/ST2/00034 (J.Z.) and Unisono 2017/25/Z/ST2/03029 (T.C.) realized within QuantERA ERA-NET QTFLAG collaboration.
\end{acknowledgments}


\appendix

\section{Details on the DMRG algorithm} \label{App:DMRG}
The phase diagrams are calculated by means of a DMRG numerical program using the ITensor C++ library \cite{itensor} (see also Ref. \cite{Kraus2020})) and 
using infinite DMRG (iDMRG) method available in TeNPy library \cite{tenpy}. 

\subsection{DMRG for finite chains}

For the finite chain we lift the degeneracy in the CDW phase and Haldane phase by adding the boundary term
$\hat{H}_{ad}=[2\rho]\left(V\hat{n}_1+V_{\text{NNN}}\hat{n}_2\right)$. The maximal bond dimension is set to $\beta = 600$, the energy error goal is fixed to $\epsilon_{goal}=10^{-16}$ and the upper limit $\epsilon$ for the singular values discarded is set to  $\epsilon = 10^{-16}$.
We allow for maximally $n_{\text{max}}=10$ particles per site.  In order to ensure that the simulations end up in the ground state we run the simulation for three different initial states: 
the CDW state 
\begin{align}
\ket{\Phi}_{\text{init}}=\otimes_{k} \ket{2\cdot\rho}_k \otimes_{l} \ket{0}_l \label{CDW_state}
\end{align} with $k \in \left\lbrace \mathbb{A}={2\cdot m|m\in \mathbb{N}} \right\rbrace$ and $l \in \mathbb{N}\backslash \mathbb{A}$, the MI state $\ket{\Phi}_{\text{init}}=\otimes_{k=1}^L\ket{\rho}_k$ and a random initial state. The random state is a superposition of Fock states $\ket{\Phi}_{\text{init}}=\frac{1}{\sqrt{n_{\text{iter}}}}\sum^{n_{\text{iter}}}_k \left(\otimes_i\ket{n_i}\right)_k$, where $n_i \in \mathbb{N}$ is chosen randomly out of the interval $[0,n_{\text{max}}]$ with the constrain $\sum_{i=1}^Ln_i=\rho$. We choose the number of superimposed Fock state to be $n_{\text{iter}}=100$. \new{We note that the string order parameter $\mathcal{O}_S(\rho)$, Eq. (\ref{eq:string}), and the structure form factor, Eq. (\ref{DW_orderparameter}), at $k=\pi$ have the same value for the CDW Fock state (see Eq. (\ref{CDW_state})) modulo a term proportional to $1/L$ and which vanishes in the limit $L\rightarrow \infty$.}
In order to calculate the first excited state one adds an extra term to the Hamiltonian, which lifts the energy of the ground state:
\begin{align}
\hat{H}_{BH}' = \hat{H}_{BH}+W\ket{\psi_0}\bra{\psi_0} \label{Hweight}
\end{align} 
 with $\ket{\psi_0}$ the ground state. The first excited state is determined by calculating  the ground state of $\hat{H}_{BH}'$ in Eq. (\ref{Hweight}) using the DMRG ground state algorithm. The weight of the extra term is chosen to be $W= 20t$. 

{We determine the ground state by means of this DMRG numerical program and calculate the observables presented in Sec. \ref{Sec:2.1}. \new{In order to get rid of the boundary effect we neglect the outer $L/4$ sites in the determination of the observables \new{following} \cite{Rossini2012,batrouni2014competing}. \new{To} justify this cut we show in Fig.~\ref{cut} the string order parameter, $\mathcal{O}_S(\rho)$~\eqref{eq:string}, as a function of the number of lattice sites cut at the boundary together with the value of the order parameter calculated by means of iDMRG. \new{For a}  systematic analysis of the effect of the boundary conditions \new{see} \cite{stumper2020macroscopic}.}

\begin{figure}[h]
	\centering
	\includegraphics[width=0.8\linewidth]{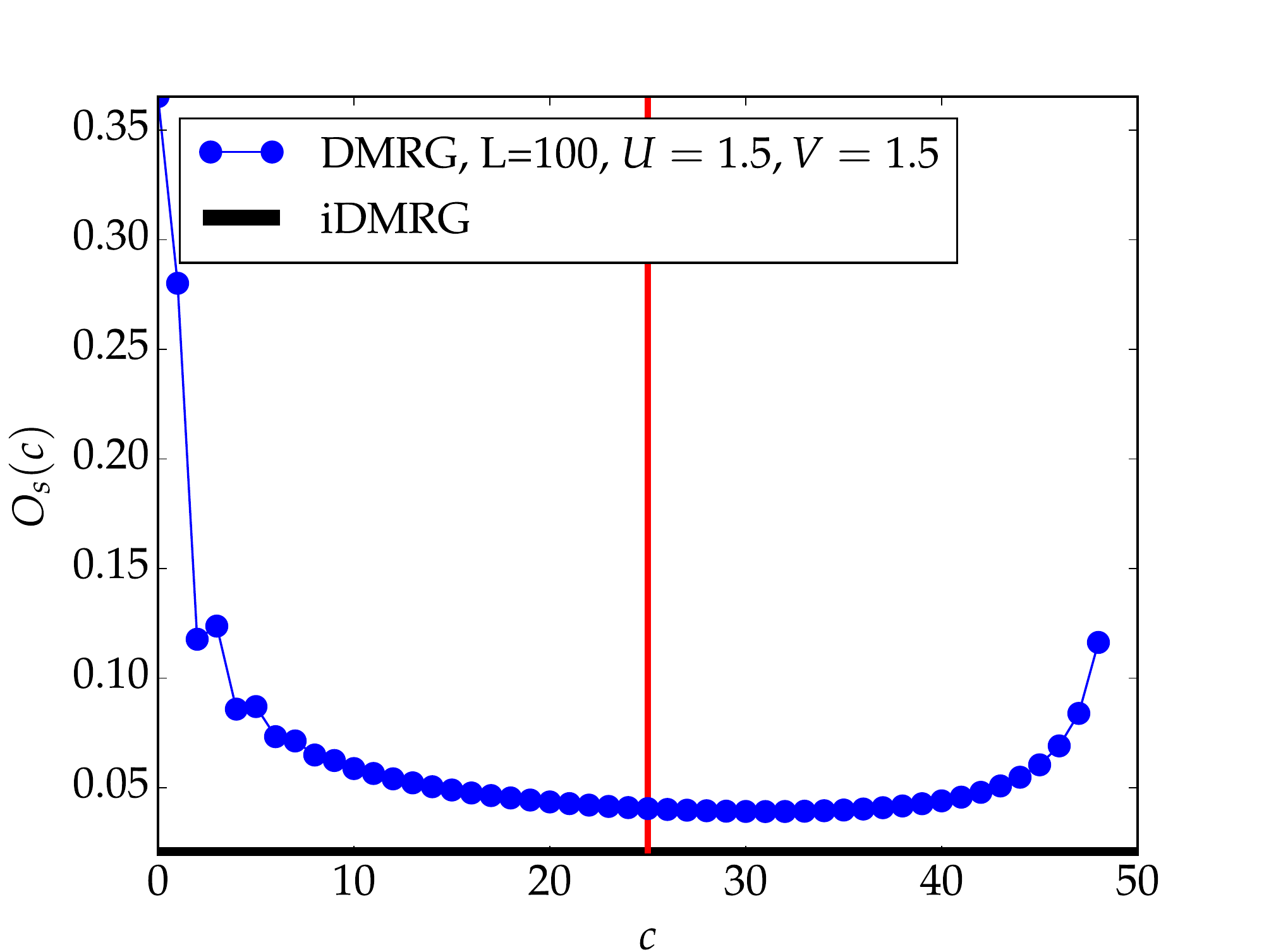} \\
	\includegraphics[width=0.8\linewidth]{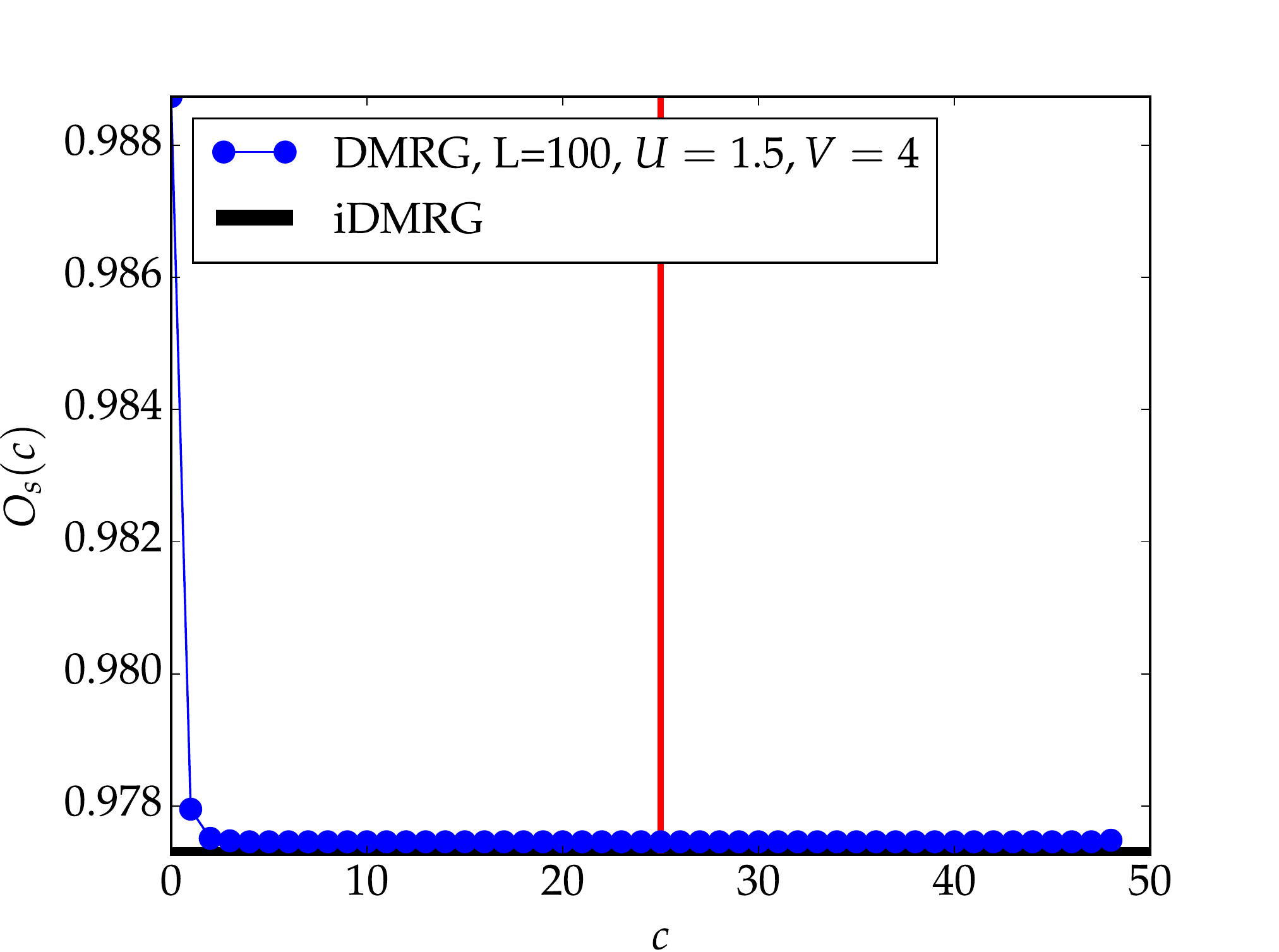}
	\caption{\new{String order parameter $\mathcal{O}_S(\rho)$, \eqref{eq:string}, (blue dots) as function of the number of cutted lattice sites $c=i$ with $i+r=L-c$ for $L=100$ and compared to the value of the string order parameter calculated by means of iDMRG (black line). The red line indicates the value at which we cut the boundaries to produce the phase diagrams in the main text. The upper subplot shows the string order parameter within the HI phase for $U=1.5$ and $V=1.5$ and the lower one shows the string order parameter within the CDW phase for $U=1.5$ and $V=4$. }} 
	\label{cut}
\end{figure}
	
\new{To get the phase diagram in the thermodynamic limit we fit the values of the observables for different numbers of lattice sites according to \eqref{O:L}. We justify the application of  Eq.~\eqref{O:L} by inspecting the observables as a function of $1/L$. Fig.~\ref{finite_size_sc} shows the neutral gap (\ref{neutral_gap}), the charge gap (\ref{charge_gap}), and the parity order parameter (\ref{eq:parity}) as a function of $1/L$ near the SF-MI phase transition at $(U/t,V/t)=(2.5,0.8)$.
The gaps follow a linear behavior as function of $1/L$ near the SF-MI transition. 
Moreover, Fig.~\ref{finite_size_sc} shows the behavior of the observables as a function of $1/L$ at the HI-MI and MI-CDW transition, where the observables \new{$1/L$ dependence is nicely fitted by (\ref{O:L}). }
\begin{figure*}[htb]
	\centering
	\includegraphics[width=0.8\linewidth]{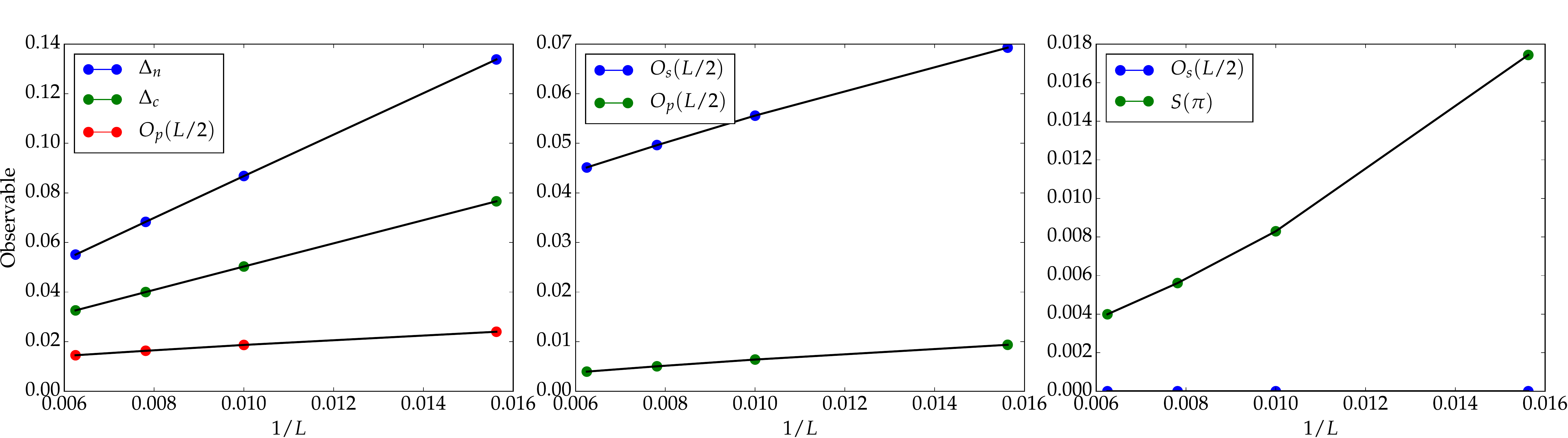} 
	\caption{\new{Values of the observables (see inset) as a function of $1/L$ around at the SF-MI transition at $(U/t,V/t)=(2.5,0.8)$ (left), the HI-MI transition $(U/t,V/t)=(3,1.85)$ (middle) and the MI-CDW transition $(U/t,V/t)=(8,4.05)$ (right). The dots show the values of the observables, whereas the black lines depict the correspond fit according to Eq.~\eqref{O:L}.}} 
	\label{finite_size_sc}
\end{figure*}	
 \new{We identified the boundary lines in Figures \ref{fig1}, \ref{String_inf}, \ref{vonNeu} and \ref{Fig:6a} by using a certain threshold value for the order parameter above which we determine a certain phase. Those threshold values are those, who reproduce the critical value of the MI-SF transition at $V/t=0$ in Ref. \cite{kuhner2000one} and the SF-HI transition at $U/t=2$ in Ref. \cite{batrouni2014competing}. Here we make use of  our data set for $T=0$.} We then convert the error corresponding to the fitting procedure into an error in the phase boundary. Fig. \ref{finite_size_sc} displays the phase boundaries in the $(U/t,V/t)$-plane including the error bars for each point at the phase boundary.}  
\begin{figure}[htb]
	\centering
	\includegraphics[width=0.8\linewidth]{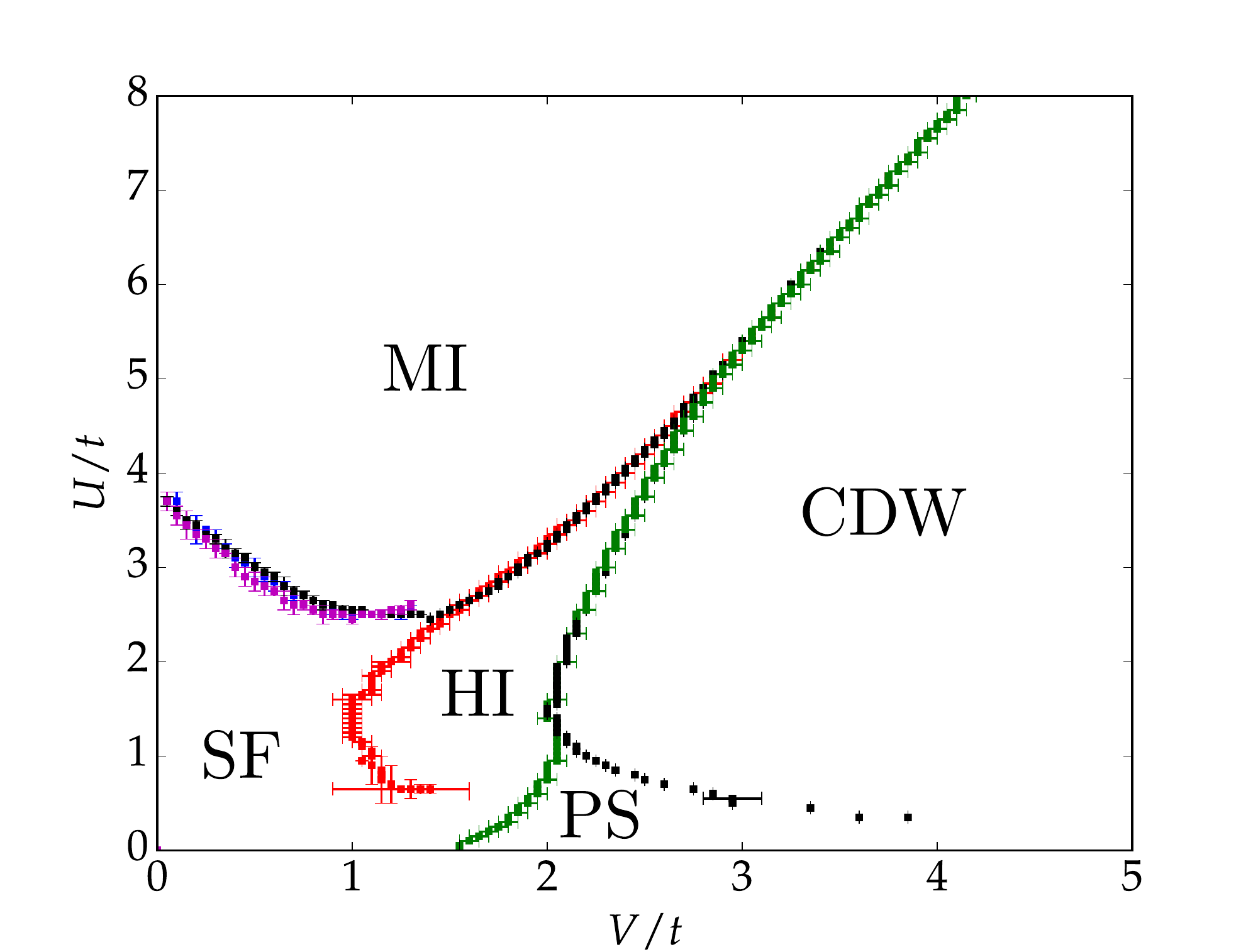}
	\caption{Same as in Fig. \ref{fig:2}, but now the error bars are explicitly shown for every reported point.} 
	\label{full_error}
\end{figure}

\subsection{iDMRG simulations}
We also explore the system directly at the thermodynamic limit using the infinite DMRG  (iDMRG) algorithm \cite{McCulloch_2008,Crosswhite_2008, Kjall_PRB_2013} based on translationally invariant infinite matrix-product state (iMPS) ansatz \cite{Vidal_2007}. Since the onset of the CDW phase requires unit cells of size integer multiple of 2, we consider iMPS representation with unit cells of size 4 for our simulations. The maximum bosonic occupancy is taken to be $n_{\text{max}}=8$. \new{We fix the maximal iMPS bond dimension to $\beta = 640$ and checked that our results do not change by changing the bond dimension to $\beta = 384, 512$.} To confirm the convergence of the iDMRG algorithm, we follow the change in energy  density in successive iDMRG sweeps, and when the change falls below $10^{-12}$, we conclude that the resulting iMPS is the ground state of the infinite system.

\newpage
\input{Haldane_arx.bbl}

\end{document}

%% file: Haldane_arx.bbl
%